\documentclass[aps,amssymb,amsmath,reprint,noshowpacs,superscriptaddress,floatfix]{revtex4-1}

\usepackage{times}
\usepackage{amssymb,amsmath,graphicx}
\usepackage[usenames]{color}
\usepackage{bbold}
\usepackage[english]{babel}
\usepackage{xr}
\usepackage{dsfont}
\usepackage{dcolumn} 
\usepackage{scalefnt}   
\usepackage{hyperref}	
\usepackage[capitalize]{cleveref}
\crefname{appsec}{Appendix}{Appendices}

\newcommand{\figwidth}{1.0\columnwidth} 
\newcommand{\commentOut}[1]{}

\newcommand{\ooc}{\omega_\mathrm{c}}
\newcommand{\oo}{\omega}
\newcommand{\oz}{\omega_0}

\newcommand{\OO}{\Omega}

\newcommand{\kk}{\kappa}

\newcommand{\kr}{\kappa_\mathrm{r}}

\newcommand{\kabs}{\kappa_\mathrm{abs}}
\newcommand{\gi}{\gamma_\mathrm{i}}
\newcommand{\gr}{\gamma_\mathrm{r}}

\newcommand{\Gr}{\tensor{G}}

\newcommand{\Gbg}{G_\mathrm{bg}}

\newcommand{\E}{\mathbf{E}}

\newcommand{\Ec}{E_\mathrm{c}}
\newcommand{\Em}{\mathbf{E}_{m}}

\newcommand{\Etot}{E_\mathrm{tot}}

\newcommand{\Ecdr}{E_\mathrm{c,drive}}
\newcommand{\Epdr}{E_{p, \mathrm{drive}}}
\newcommand{\ac}{a_\mathrm{c}}
\renewcommand{\r}{\mathbf{r}}
\newcommand{\ro}{\mathbf{r}_0}

\newcommand{\ee}{\epsilon}
\newcommand{\eo}{\epsilon_0}

\newcommand{\rz}{\mathbf{r}_0}
\newcommand{\rdr}{\mathbf{r}_\mathrm{dr}}
\renewcommand{\Re}[1]{\ensuremath{\operatorname{Re}\left\{#1\right\}}}
\renewcommand{\Im}[1]{\ensuremath{\operatorname{Im}\left\{#1\right\}}}
\newcommand{\Pdr}{P_{\mathrm{dr}}}
\newcommand{\PdrSE}{P_{\mathrm{dr,SE}}}
\newcommand{\ettot}{\eta_\mathrm{tot}}
\newcommand{\etptot}{\eta_{p, \mathrm{tot}}}
\newcommand{\etctot}{\eta_{c, \mathrm{tot}}}

\newcommand{\etabs}{\eta_{p, \mathrm{abs}}}
\newcommand{\etprad}{\eta_{p, \mathrm{rad}}}
\newcommand{\etcrad}{\eta_\mathrm{c,rad}}

\newcommand{\etcabs}{\eta_\mathrm{c,abs}}

\newcommand{\etlim}{\eta_\mathrm{ant}^\mathrm{lim}}
\newcommand{\etSE}{\eta_\mathrm{SE}}
\newcommand{\p}{\mathbf{p}}
\newcommand{\pdr}{p_{\mathrm{dr}}}
\newcommand{\ptot}{p_{\mathrm{tot}}}
\newcommand{\pSE}{p_\mathrm{SE}}
\newcommand{\rbg}{\rho_\mathrm{bg}}

\newcommand{\ahom}{\alpha_\mathrm{hom}}
\newcommand{\ahyb}{\alpha_\mathrm{H}}
\newcommand{\alim}{\alpha^\mathrm{lim}}

\newcommand{\Veff}{V_\mathrm{eff}}
\newcommand{\Veffhyb}{V_\mathrm{eff}^{\mathrm{hyb}}}
\newcommand{\Um}{U_\mathrm{m}}

\newcommand{\chihom}{\chi_\mathrm{hom}}
\newcommand{\chihyb}{\chi_\mathrm{H}}

\newcommand{\hatp}{\mathbf{\hat{p}}}

\newcommand{\Fp}{F_{\mathrm{P}}}

\newcommand{\ii}{\mathrm{i}}

\begin{document}

\title{Antenna-cavity hybrids: matching polar opposites for Purcell enhancements at any linewidth}
\author{Hugo M. Doeleman}
\email{h.doeleman@amolf.nl}
\affiliation{Center for Nanophotonics, FOM Institute AMOLF,
Science Park 104, 1098 XG Amsterdam, The Netherlands}
\affiliation{Van der Waals-Zeeman Institute, University of Amsterdam, Science Park 904, PO Box 94485, 1090 GL Amsterdam, The Netherlands}
\author{Ewold Verhagen}
\affiliation{Center for Nanophotonics, FOM Institute AMOLF,
Science Park 104, 1098 XG Amsterdam, The Netherlands}
\author{A. Femius Koenderink}
\affiliation{Center for Nanophotonics, FOM Institute AMOLF,
Science Park 104, 1098 XG Amsterdam, The Netherlands}
\affiliation{Van der Waals-Zeeman Institute, University of Amsterdam, 
Science Park 904, PO Box 94485, 1090 GL Amsterdam, The Netherlands}

\begin{abstract}
Strong interaction between light and a single quantum emitter is essential to a great number of applications, including single photon sources. Microcavities and plasmonic antennas have been used frequently to enhance these interactions through the Purcell effect. Both can provide large emission enhancements: the cavity typically through long photon lifetimes (high $Q$), and the antenna mostly through strong field enhancement (low mode volume $V$). In this work, we demonstrate that a hybrid system, which combines a cavity and a dipolar antenna, can achieve stronger emission enhancements than the cavity or antenna alone. We show that such systems can be used as a versatile platform to tune the bandwidth of enhancement to any desired value, while simultaneously boosting emission enhancement. Our fully consistent analytical model allows to identify the underlying mechanisms of boosted emission enhancement in hybrid systems, which include radiation damping and constructive interference between multiple-scattering paths. Additionally, we find excellent agreement between strongly boosted enhancement spectra from our analytical model and from finite-element simulations on a realistic cavity-antenna system. Finally, we demonstrate that hybrid systems can simultaneously boost emission enhancement and maintain a near-unity outcoupling efficiency into a single cavity decay channel, such as a waveguide.
\end{abstract}
\date\today

\maketitle


\section{Introduction}
For many nanophotonic applications, such as single photon sources operated at high frequency \cite{Lounis2005,Eisaman2011,Lodahl2015}, nanoscale lasers \cite{Hill2014}, quantum logical gates for photons \cite{Kimble2008,OBrien2009} and highly sensitive, low detection volume sensing devices \cite{Anker2008,Qavi2009,Vollmer2012}, strong interactions between a single quantum emitter and light are vital. 
This interaction can be enhanced by coupling emitters to nanophotonic structures that enhance the emission rate of the emitters using the Purcell effect \cite{Purcell1946}. 
Tradionally, enhanced emission rates are achieved by placing emitters in dielectric microcavities. 
The relative emission enhancement of an emitter at resonance with a cavity mode, i.e. the Purcell factor ($\Fp$), then relates to the quality factor ($Q$) and the mode volume ($V$) as
\begin{equation}
\Fp= \left( 3/(4\pi^2) \right) \left( \lambda / n \right)^3 \left( Q / V \right), \label{Eq:Purcell}
\end{equation}
where $n$ is the index of the medium around the emitter. Microcavity modes typically reach large enhancements because of their extremely long photon lifetimes and consequently high quality factors \cite{Vahala2003}. Additionally, almost all of the light is typically emitted into a single photonic mode, facilitating efficient collection through e.g. a waveguide, which is a major advantage for applications such as single photon sources \cite{Arcari2014,Lodahl2015}.
Plasmonic nano-antennas are a popular alternative solution \cite{Novotny2011,Tame2013fixed}. Rather than storing photons for a very long time, antennas are able to concentrate their energy in volumes far below the diffraction limit \cite{Gramotnev2010,Schuller2010}, thus achieving unparalleled emission enhancements over large bandwidths \cite{Akselrod2014}. 

Both microcavities and antennas also suffer from important drawbacks. Microcavities are limited in their mode volume by the diffraction limit, hence requiring high quality factors to compensate. Unfortunately, a high $Q$ can be unpractical for several applications. For instance, high-$Q$ cavities are often extremely sensitive to changes in temperature and environment, as well as to minor fabrication errors, making it difficult to scale to multiple connected devices in e.g. a quantum photonic network \cite{Kimble2008,OBrien2009}. Moreover, such narrow resonances typically do not match with the broad emission spectra of room temperature single-photon emitters.
Antennas, on the other hand, suffer from strong radiative and dissipative losses, which limit $Q$ to $\sim$10-50. This makes applications in quantum information difficult, because it would require emitter-antenna strong coupling, i.e. coupling rates higher than the antenna loss rate \cite{Kimble1998,Raimond2001}. Also, their non-directional emission patterns tend to make efficient collections of the emission difficult.
Ideally, one would be free to choose any desired $Q$, independent of the Purcell factor. An attractive candidate to achieve such tunability is a hybrid cavity-antenna system.
Recently such systems were proposed for a selection of applications including emission enhancement \cite{Barth2010,Boriskina2011}, molecule or nanoparticle detection \cite{DeAngelis2008,Santiago-Cordoba2012,Hu2013,Conteduca2015}, nano-scale lasers \cite{Zhang2014a,Mivelle2014} and strong concentration near an antenna of light from free space or a waveguide \cite{Chamanzar2011,Ahn2012,Ren2015,Ahn2016}. Also 2D Fabry-P\`{e}rot etalons coupled to antennas have been used to study antenna-cavity coupling mechanisms \cite{Ameling2010,Vazquez-Guardado2014,Bahramipanah2015}. 

In this work, we demonstrate that hybrid cavity-antenna systems can yield emission enhancements larger than that of both the cavity and the antenna, and reveal the mechanisms behind these enhancements. 
It was predicted in earlier work that an emitter coupled to a high-$Q$ cavity could gain in emission enhancement through the inclusion of a small nano-particle \cite{Xiao2012}. However, these results did not extend to larger, strongly scattering particles because radiative antenna damping was not taken into account. In other work it was demonstrated that radiation damping can in fact be important \cite{Frimmer2012}. In fact, it was argued in \cite{Frimmer2012} that for strongly radiatively damped antennas near resonance, the predictions of \cite{Xiao2012} are completely reversed: rather than an increase, the authors found a strong \emph{suppression} of emission enhancement when the antenna was coupled to a cavity. Here, we develop a simple coupled harmonic oscillator model that includes all cavity-antenna interactions and the radiative antenna losses. This model thus holds for both weakly and strongly scattering particles, and is in fact general to all types and geometries of cavities and antennas. We show that improved enhancements in these systems result from a trade-off between additional losses and confinement, and we elucidate under what conditions one can profit maximally from these effects. We demonstrate for a wide range of cavities that hybrid systems allow to tune the bandwidth of emission --- often up to several orders of magnitude increase --- while maintaining comparable or even higher emission enhancement than the bare cavity. Moreover, we propose a realistic design for a hybrid system that can be fabricated lithographically, and validate using COMSOL simulations that our model correctly predicts the strongly increased Purcell enhancements in this design. Finally, we demonstrate that hybrid systems can boost emission enhancements while retaining a high power outcoupling efficiency into a single cavity decay channel (e.g. a waveguide), making them excellent candidates for single photon sources. 

\section{\label{sec:Model}Modelling hybrid emission enhancements}

The emission enhancement experienced by a quantum emitter due to its environment can be found by modelling the emitter as a classical oscillating dipole with constant current amplitude. The power emitted by such a drive dipole is equal to the work done by its own field on itself, i.e. \footnote{Note that all quantities are scalars, as we have projected the fields on the axis of the antenna dipole moment, and assumed the drive dipole orientation to be aligned with this axis.} 
\begin{equation}
	P_{\mathrm{dr}} = \frac{\omega}{2} \Im{\pdr^{*} \, \Etot }, 
    \label{Eq:Pdr}	
\end{equation}
where $\pdr$ is its dipole moment, $\oo$ its angular oscillation frequency and $\Etot$ the total field at its position. Dividing $P_{\mathrm{dr}}$ by the power that the drive dipole emits in a homogeneous medium, as given by Larmor's formula, yields the emission enhancement $\eta$, also known as the `local density of optical states' (LDOS) relative to the medium \cite{Novotny2012}. In the context of cavities, $\eta$ evaluated at the cavity resonance is the Purcell factor. Thus the task of finding emission enhancements reduces to that of finding the field $\Etot$.

\begin{figure}[t]
\includegraphics[width=\figwidth]{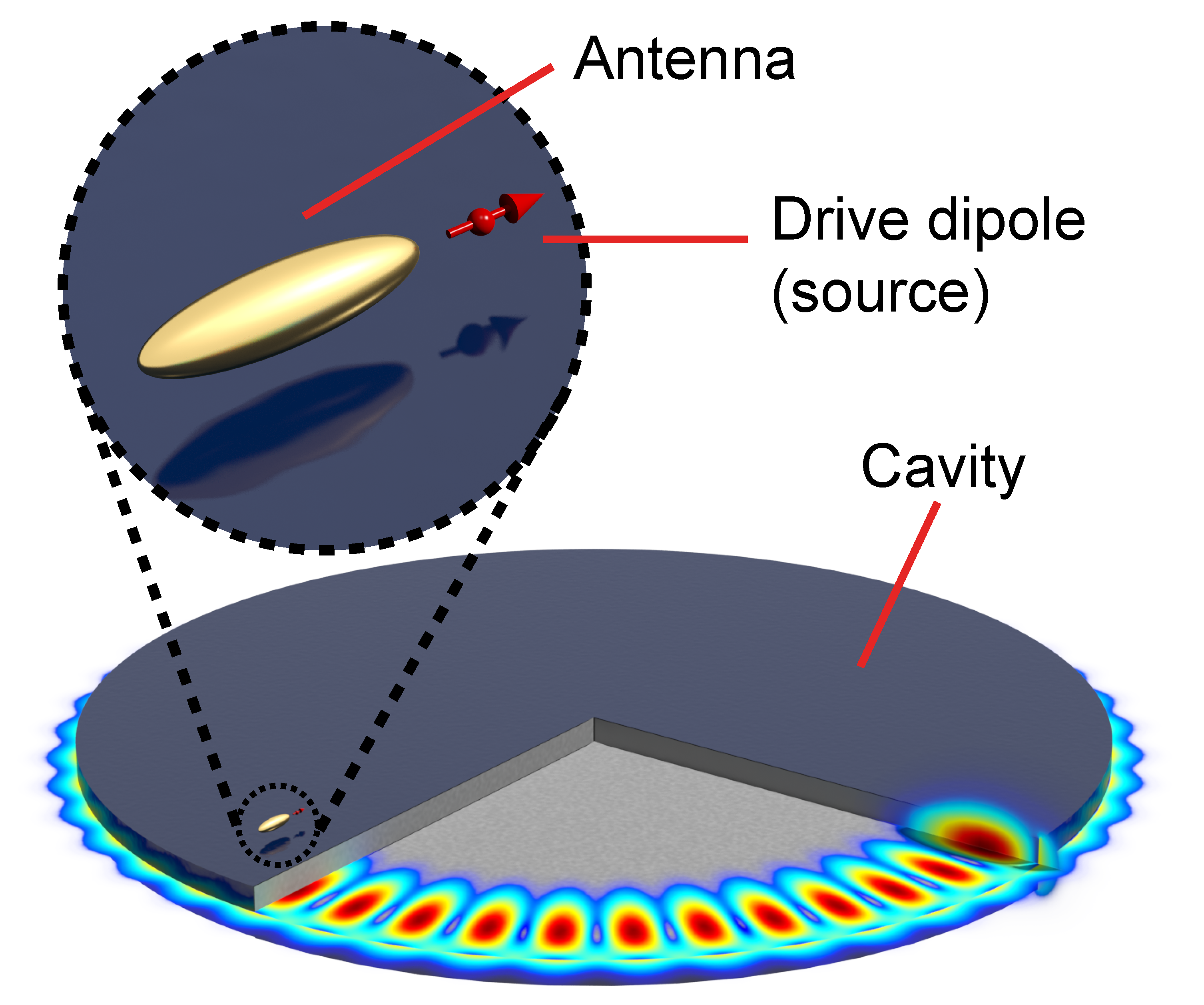} 
\caption{A coupled cavity-antenna system, driven by a dipolar source. The cavity is represented by a disk supporting a high quality factor whispering gallery mode (WGM) shown in the cut-out. The antenna is a gold ellipsoid placed in the near-field of the cavity. The source is an oscillating point dipole.} \label{fig:Cartoon} 
\end{figure}

Here we consider an emitter coupled to a cavity-antenna system. A possible configuration is depicted in \cref{fig:Cartoon}. However, no assumptions on either cavity or antenna geometry are made, other than that the antenna is dipolar. 
To obtain $\Etot$, we model cavity and antenna as harmonic oscillators and set up their coupled equations of motion (EOM). We obtain (see \cref{sec:SI-EOMS})
\begin{align}
	(\oz^2- \oo^2-i\oo \gamma ) \, p - \beta \Ec = \beta \Epdr, \label{Eq:EOMone}\\
	-\frac{\omega^2}{\eo \ee \Veff} \,p \, + (\ooc^2- \oo^2-i\oo \kk )\, \Ec 
	= \oo^2 \Ecdr , \label{Eq:EOMtwo}
\end{align}
where the free variables $p$ and $\Ec$ are the antenna induced dipole moment and cavity mode field amplitude at the position of the antenna, respectively. It is easy to see that \cref{Eq:EOMone} exactly maps on a point dipole model for a polarizable plasmon antenna, driven by an external driving field $\Epdr$ and the cavity field \cite{Lagendijk1996}. Likewise, in \cref{Eq:EOMtwo} one recognizes the typical description of the response of a single cavity mode, driven by an external field $\Ecdr$ and the antenna. Antenna and cavity resonance frequencies are denoted by $\oz$ and $\ooc$, respectively, and their respective damping rates by $\gamma$ and $\kappa$. Importantly, $\gamma$ contains an intrinsic damping rate $\gi$ due to ohmic damping, and a frequency-dependent radiative damping rate $\gr$, through 
\begin{equation}
\gamma (\oo)= \gi +\gr (\oo).
\end{equation}
The radiative damping rate $\gr$ is proportional to the LDOS of the background environment, i.e. including all optical modes at the position of the antenna, yet \emph{excluding} the cavity mode under consideration. The antenna will also experience additional radiation damping due to the cavity, which is separately accounted for through the EOM, as will become apparent below. The antenna-cavity coupling is determined by the antenna oscillator strength $\beta$ and the bare cavity effective mode volume $\Veff$. 
While in a Drude model for a metal sphere of volume $V_\mathrm{ant}$ in vacuum, $\beta$ simply reads $3 V_\mathrm{ant} \eo \oz^2$, in general it may be found for any antenna by polarizability tensor retrieval from a full wave simulation \cite{BernalArango2013,BernalArango2014}. 
The effective mode volume $\Veff$ equals the mode volume $V$ in \cref{Eq:Purcell} if the antenna is placed exactly at the cavity mode maximum. Away from the mode maximum, $\Veff$ increases, in inverse proportion to the mode profile (see \cref{Eq:SI-Veff} for an exact definition). Finally, the relative permittivity $\ee$ refers to the medium surrounding the antenna. 

We may now identify $\Epdr$ and $\Ecdr$ with the field generated by the drive dipole as $\Epdr = \Gbg \, \pdr $ and $ \Ecdr = \pdr / \left( \eo \ee \Veff \right) $. Note that, in contrast to the induced dipole moment $p$ of the antenna, the drive dipole has fixed dipole moment $\pdr$. The Green's function of the background environment $\Gbg =\hatp \cdot \Gr_\mathrm{bg} (\rdr , \rz , \omega) \cdot \hatp_\mathrm{dr}$ describes the field caused by the drive dipole, at the position of the antenna. The same effective mode volume $\Veff$ as in \cref{Eq:EOMtwo} appears here, if we assume the cavity mode fields at the positions of the drive dipole and antenna to be equal. This is true if the distance between them is much smaller than the wavelength. If this is not the case, our formalism remains applicable, however one should include a complex factor in $\Ecdr$ to account for the difference in amplitude and phase of the cavity mode field at the drive dipole position and the antenna position. 

If we consider first the uncoupled EOMs, we can recognize the bare antenna polarizability $\ahom$ and bare cavity response $\chihom$, defined through $p=\ahom \Epdr$ and $\Ec=\chihom \pdr$, respectively. These are
$\ahom= \beta / (\oz^2- \oo^2-i\oo \gamma )$ and 
$\chihom = \left(\oo^2 / \eo \ee \Veff \right) / \left( \ooc^2- \oo^2-i\oo \kk \right)$.
When cavity and antenna are coupled, their own scattered fields act as additional driving terms. These fields can be expressed in an infinite series of cavity-antenna interactions, similar to a multiple-scattering series in a coupled point-scatterer model \cite{Lagendijk1996,GarciadeAbajo2007}. The series can be captured in the hybridized antenna polarizability $\ahyb$ and cavity response function $\chihyb$, which account for all possible interactions between cavity and antenna. They are given as
\begin{align}
\ahyb&= \ahom \left( 1 - \ahom \chihom \right)^{-1}, \label{Eq:ahyb}\\
\chihyb &= \chihom \left( 1 - \ahom \chihom \right)^{-1}. \label{Eq:chihyb}
\end{align}
The hybridized polarizability $\ahyb$ resembles the broad, Lorentzian lineshape of $\ahom$, yet with a sharp Fano-type resonance close to $\ooc$, similar to the polarizability discussed by Frimmer et al. \cite{Frimmer2012}. Increased radiation damping experienced by the antenna due to the cavity mode, as measured by Buchler et al. for a dipole near a mirror \cite{Buchler2005} , is also captured in $\ahyb$. The hybridized cavity response $\chihyb$, on the other hand, shows a Lorentzian lineshape with a resonance that is shifted and broadened exactly as predicted by the familiar Bethe-Schwinger perturbation theory \cite{Bethe1943,Waldron1960,Ruesink2015} (see \cref{sec:SI-AlphaChi}).

We can now find the total field at the drive dipole position as the sum of the cavity field, the antenna scattering and the contribution of the background medium, i.e. $E_{\mathrm{tot}}=\Ec + \Gbg p + \Gbg(\rdr, \rdr, \omega) \pdr$. Solving the EOMs yiels $\Ec$ and $p$. Using this field in \cref{Eq:Pdr} and dividing by Larmor's formula, we obtain the emission enhancement (see \cref{sec:SI-ettot}):
\begin{align}
\ettot &= 1+ \frac{6 \pi \eo c^3 }{\oo^3 n } \Im{\ahyb \Gbg^2  + 2 \Gbg \ahyb \chihom +\chihyb }. \label{Eq:ettot}
\end{align}
Note that each of the terms in $\ettot$ corresponds to a multiple scattering path that radiation can take, departing from and returning to the source, which we discuss in \cref{sec:SplitByPath}. 

\section{\label{sec:BareVsHyb}Enhancement in hybrids and bare components, and the `superemitter'}

\begin{figure}[t]
\includegraphics[width=\figwidth]{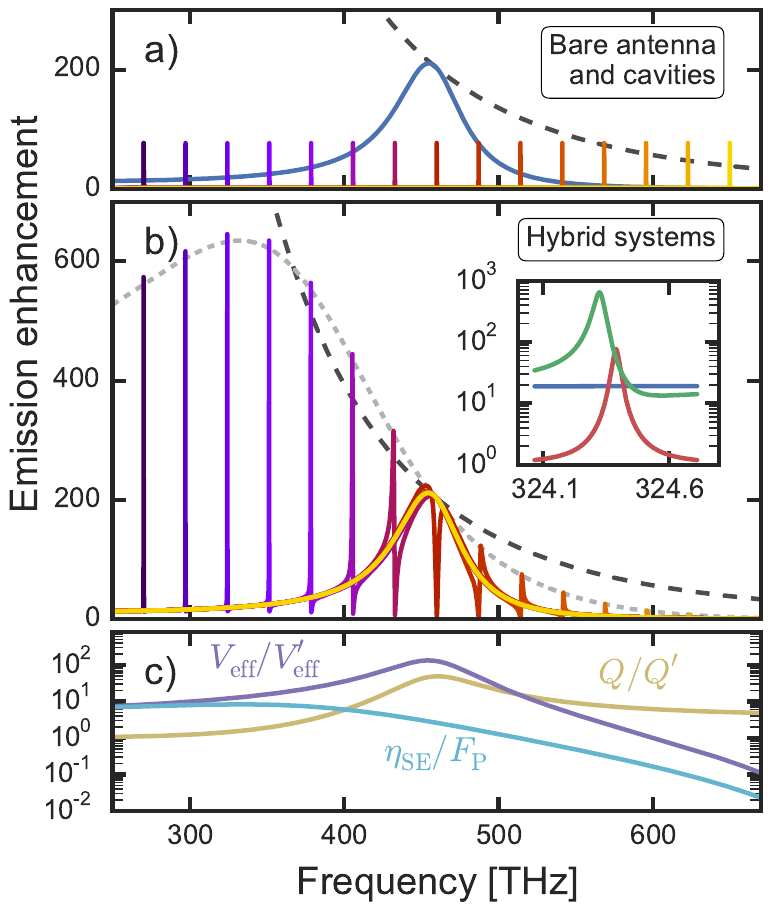} 
\caption{
\textbf{a)} Emission enhancement for a dipole coupled to a bare antenna (blue line) or to a set of bare cavity modes (other colors). Cavity resonances are spaced half an antenna linewidth (i.e. 27.1 THz) from each other. Each cavity peak represents a different calculation, indicated by a different color. The antenna limit $\etlim$ discussed in \cref{sec:SplitByPath} is shown by the dashed dark grey line. 
\textbf{b)} Emission enhancement for the hybrid system (colored lines) composed of the same elements as shown in Fig. a, compared to $\etlim$ (dashed dark grey line). Each spectrum contains both a narrow Fano-resonance and a broad peak near the antenna resonance frequency. The peak enhancement $\etSE$ derived from a superemitter approximation (light grey dashed line) shows good agreement with the narrow peaks away from the antenna resonance. The inset contains a zoom-in on the peak with highest emission enhancement, showing antenna (blue), cavity (red) and hybrid (green) enhancements. 
\textbf{c)} Broadening (yellow) and confinement (purple) of the hybrid system, approximated as a superemitter, relative to the bare cavity. The cyan line shows the ratio of the confinement and the broadening, which equals the peak enhancement of the superemitter $\etSE$ relative to the bare cavity Purcell factor $\Fp$.} \label{fig:EnhUL} 
\end{figure} 

Using \cref{Eq:ettot}, we may now compare hybrid enhancements with those in the bare cavity and antenna. For concreteness we focus on a particular example cavity and antenna, for which \cref{fig:EnhUL}a shows enhancement spectra. Expressions for bare component enhancements can be easily derived from \cref{Eq:ettot} by taking respectively $\beta \rightarrow 0$ or $\Veff \rightarrow \infty$ (see \cref{sec:SI-ettot}). For the antenna, we take $\beta =$0.12 C$^2$/kg, corresponding to a 50 nm radius sphere with resonance frequency $\oz / (2 \pi ) = 460$ THz, and an ohmic damping rate $\gamma /(2 \pi ) = 19.9$ THz corresponding to that of gold \cite{Penninkhof2008}. The antenna is assumed to be in vacuum. We place the source at 60 nm distance from the antenna center, chosen such that we can safely neglect quenching by modes other than the dipolar \cite{Mertens2009}, with its dipole moment pointing away from the antenna. This yields an emission enhancement of about 200 at resonance. For the cavity we assume $Q \equiv \ooc/\kk =10^4$ and an effective mode volume of 10 cubic wavelengths ($\lambda$), leading to a cavity Purcell factor of 76. We present results for several different cavity resonance frequencies $\ooc$. Note that we could also have chosen to keep $\ooc$ constant and vary $\oz$ instead. 

\cref{fig:EnhUL}b shows the enhancement spectra for the hybrid systems composed of the aforementioned components. Each hybrid system has a spectrum containing 2 features, corresponding to the two eigenmodes of the system: a broad resonance peak due to a mode very similar to the bare antenna resonance, and a narrow resonance near the bare cavity resonance frequency, which originates from the 'cavity-like' eigenmode. See \cref{fig:SI-Hybridspectrum} for an example of a single hybrid spectrum. In the remainder of this paper, we will focus only on the narrow resonance. Because the source excites both hybrid eigenmodes, we observe a distinct Fano-type lineshape for the narrow resonance. Importantly, these Fano-resonances show peak enhancements that can far exceed those of the bare components. The hybrid system outperforms the antenna \emph{at resonance} by more than a factor 3, and the cavity by more than a factor 8. At the same detuned frequency, the antenna can be outperformed by up to a factor 25 for the lowest frequency peaks shown. Quantitatively similar behaviour was also predicted in earlier work \cite{Xiao2012}. Contrary to intuition, however, the strongest enhancements are not found for a cavity and an antenna tuned to resonance, but rather for cavities detuned from the antenna, and in particular for significant red-detuning. 
On resonance the cavity and antenna modes destructively interfere to yield a strongly suppressed enhancement, consistent with the findings of Frimmer et al. for hybrid system with a strongly radiatively damped antenna \cite{Frimmer2012}. 

To understand why it is possible to boost emission enhancement so strongly compared to the bare components, we can employ a `superemitter' point of view. This concept was originally proposed by Farahani et al., who claimed that an emitter coupled to an antenna could be considered as one large effective dipole when interacting with its environment \cite{Farahani2005}. Hence, for a superemitter coupled to a cavity, the emitted power is given by the expression for a dipole in a cavity, i.e.
\begin{equation}
\PdrSE = \frac{\oo}{2}|\pSE|^2 \Im{\chi}, \label{Eq:Pse}
\end{equation}
where $p_\mathrm{SE}=\pdr +p=\pdr \left(1+\Gbg \alpha \right)$ is the effective dipole moment of the superemitter, $\chi$ is the cavity response and $\alpha$ is the antenna polarizability. The picture of a superemitter acting as an ordinary emitter, yet with a larger dipole moment, suggests to use both the bare polarizability $\ahom$ and the bare cavity response $\chihom$. However, Frimmer et al. demonstrated that this procedure fails to describe the dispersive Fano lineshapes and the strongly suppressed enhancement at the antenna resonance \cite{Frimmer2012}. Frimmer et al., proposed to use the hybridized polarizability $\ahyb$ paired with $\chihom$ instead, which does predict the correct lineshapes and predicts suppression at antenna resonance. A third, alternative approach would be to use $\ahom$ for the antenna, yet describe the cavity using the hybridized response $\chihyb$. Compared to the full, self-consistent expression \cref{Eq:ettot} for emission enhancements, all three superemitter descriptions are oversimplified, and only explain particular aspects of the mechanisms behind hybrid emission enhancements. The merit of using $\ahom$ and $\chihyb$ is that it accurately predicts the \emph{envelope} function encompasssing the Fano features. Indeed,  while this third approach predicts Lorentzian peaks and fails to describe the dispersive Fano-lineshapes close to antenna resonance, it serves as a good measure for the \emph{amplitude} of the peaks. In this approach, at  a hybrid  resonance the emission enhancement experienced by a drive dipole in a superemitter can then be straightforwardly derived as $\etSE= 3/(4 \pi^2) Q'/ \Veff'$, with $\Veff'=\Veff/ |1+\Gbg \ahom|^2$ a perturbed cavity mode volume (in cubic wavelengths) and $Q'\approx \ooc / \kappa'$, where $\kappa'= \kappa + \left(\ooc / \eo \ee \Veff \right) \Im{\ahom(\ooc)}$. In the second term of $\kappa'$, one recognizes the familiar result from perturbation theory, which states that a cavity resonance is broadened by the scatterer \cite{Waldron1960}. This superemitter description thus allows us to describe the emission enhancement as a balance between enhanced broadening and confinement. 

\cref{fig:EnhUL}c shows the extra confinement $\Veff / \Veff'$ and broadening $Q/Q'$ of the superemitter relative to the bare cavity. We see broadening is dominant on the blue side of the resonance, because of increased radiation damping of the antenna. The LDOS of a homogeneous medium is strongly frequency dependent, leading to an increased radiative damping rate $\gr$ for higher frequencies. Confinement, instead, favours detunings to the red of the antenna resonance. This is due firstly to the lower radiation damping, and secondly to the positive sign of $\Re{\ahom}$, which leads to constructive interference between source and antenna when radiating into the cavity. On the blue side $\Re{\ahom}$ is negative, leading to destructive interference \footnote{Note that at this small antenna-source distance, $\Gbg$ is almost entirely real over the spectrum shown in \cref{fig:EnhUL}.}. Combined, these effects cause the emission enhancement relative to the bare cavity (cyan line in \cref{fig:EnhUL}c) to be largest on the red side of the antenna resonance. 
The corresponding peak enhancement $\etSE$, which is shown by the light grey dashed curve in \cref{fig:EnhUL}b, is in good agreement with the height of the peaks from the complete model. 
Based on the expressions for $Q'$ and $\Veff'$, we speculate that confinement can be further boosted without increasing broadening using an antenna with stronger coupling to emitters. For instance, bow-tie antennas have similar dipole moments yet larger field enhancements (captured in $\Gbg$) \cite{Kinkhabwala2009}. In fact, earlier finite-element simulations on a hybrid system composed of a nanobeam cavity and a bow-tie antenna showed a reduction of the cavity mode volume, due to inclusion of the antenna, of more than a factor 1000, with only a minor effect on $Q$ \cite{Conteduca2015}.
These results show that hybrid systems, with the right choice of cavity-antenna detuning, are able to achieve the best of both worlds: a high Q-factor typical for dielectric cavities, combined with a strongly decreased mode volume due to the high field confinement by the antenna. As an example, the inset in \cref{fig:EnhUL}b shows a hybrid mode with $Q$=$6.9 \cdot 10^3$ very similar to the bare cavity ($10^4$), but mode volume decreased by an order of magnitude (from $10\lambda^3$ to $0.82\lambda^3$).


\section{\label{sec:SplitByPath}Enhancements split by radiation path}

\begin{figure}[t]
\includegraphics[width=\figwidth]{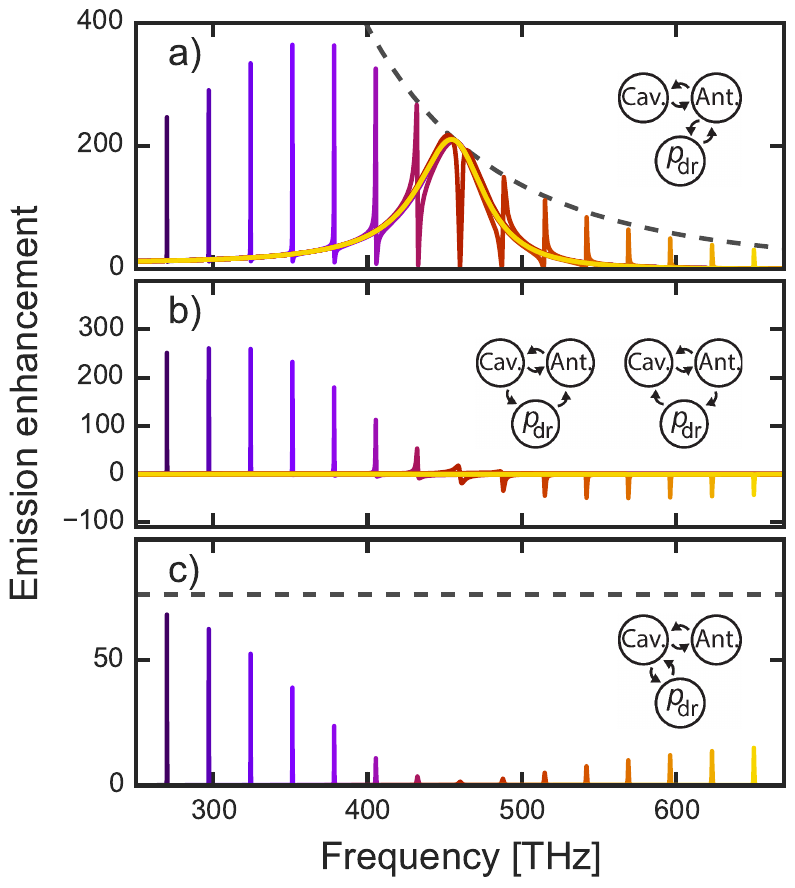} 
\caption{Emission enhancement for the hybrid system, broken down into 3 contributions corresponding to the terms in brackets in \cref{Eq:ettot}. the `antenna' term (top graph), the `cross-terms' (middle graph) and the `cavity' term (bottom graph). Each contribution corresponds to a radiation path, which are shown in the insets. The grey dotted lines in the top and bottom graphs show $\etlim$ and the bare cavity Purcell factor $\Fp$, respectively. } \label{fig:BreakdownTerms} 
\end{figure}


A different viewpoint can be obtained by analyzing \cref{Eq:ettot}, which indicates that three different multiple-scattering pathways contribute to the emission enhancement \footnote{The contribution of the fourth scattering path through the background medium, i.e. the 1 in \cref{Eq:ettot}, is trivial and not discussed here.}. We will refer to the first, second and last term in brackets in \cref{Eq:ettot} as the `antenna' term, `cross-term' and `cavity' term, respectively.  \cref{fig:BreakdownTerms} shows the hybrid enhancements from \cref{fig:EnhUL}b broken down into these three terms in \cref{Eq:ettot}, with the corresponding scattering paths shown in the insets. We see in \cref{fig:BreakdownTerms}a that the antenna term, corresponding to scattering paths that start and end with an antenna-source interaction, is dominant over most of the spectrum. However, we also recognize that this term alone cannot break the bare antenna limit, which is shown as the grey dotted line. This limit follows from the well-known upper bound of $(3/(2\pi^2))\lambda^2$ on the extinction cross section of a single dipolar scatterer, which is a consequence of energy conservation \cite{Foot2005,Ruan2011,Frimmer2012}. Consequently its polarizablity is limited to $|\alim|=\Im{\alim}=(3\eo \ee /(4\pi^3)) \lambda^3$. An antenna with an albedo $A=\gr/(\gi+\gr)$ of 1 reaches this limit at its resonance frequency. The limit on $\alpha$ leads to a limit on antenna enhancement given by
$\etlim = 1+ 6 \pi \eo c^3 / \left( \oo^3 n \right) \Im{\alim \Gbg^2 } A(\oo) $, 
where we have included the albedo to account for ohmic damping in the antenna. Not only a bare antenna, but also the antenna term in \cref{fig:BreakdownTerms}a must obey this limit, because $\ahyb$ remains bound by $\alim A(\oo)$ through energy conservation.

Despite the antenna term being bound to the antenna limit, the total enhancement $\ettot$ can break this limit. This is due to the contributions of the cavity term and the cross-term, both of which require direct interaction between cavity and source, i.e. without passing through the antenna in between. The cavity term in \cref{fig:BreakdownTerms}c, which represents all scattering paths starting and ending with a source-cavity interaction, is relatively weak. This is because the perturbed cavity response $\chihyb$ is always weaker than that of the unperturbed cavity ($\chihom$), causing enhancement to remain below the cavity Purcell factor $\Fp$. The cross-term in \cref{fig:BreakdownTerms}b, on the other hand, contributes significantly to the hybrid enhancement. This term describes two types of scattering paths: those starting at the antenna and ending at the cavity, and vice versa. This term particularly contributes on the red side of the antenna resonance $\oz$, and in fact the enhancement switches sign at $\oz$. Similar to the interference between two fields $E_1$ and $E_2$, the sign of the cross-term ($E_1 E_2^* + E_1^* E_2$) indicates constructive or negative interference. In this hybrid system, the interference takes place inside both cavity and antenna, between light radiated by the source via the antenna on the one hand, and via the cavity on the other. Indeed, in \cref{fig:EnhUL}b we see that the sum of all three terms can break the antenna limit, indicated by the dark dashed grey curve, for frequencies where this constructive interference takes place. 

In conclusion, we have seen that direct coupling between the source and the cavity mode plays a crucial role in boosting the enhancement of a hybrid system beyond the antenna limit. Although most of the hybrid enhancement comes from the antenna term, where direct source-cavity coupling plays no role, this term alone can never break the antenna limit. However, light radiated directly by the source into the cavity - and particularly its interference with the light radiated to the antenna - allow to break this limit, thus achieving even stronger emission enhancements in the hybrid system than the antenna alone could ever achieve. 

%
\section{The range of effective hybrid $Q$ and $V$}

\begin{figure}[t]
\includegraphics[width=\figwidth]{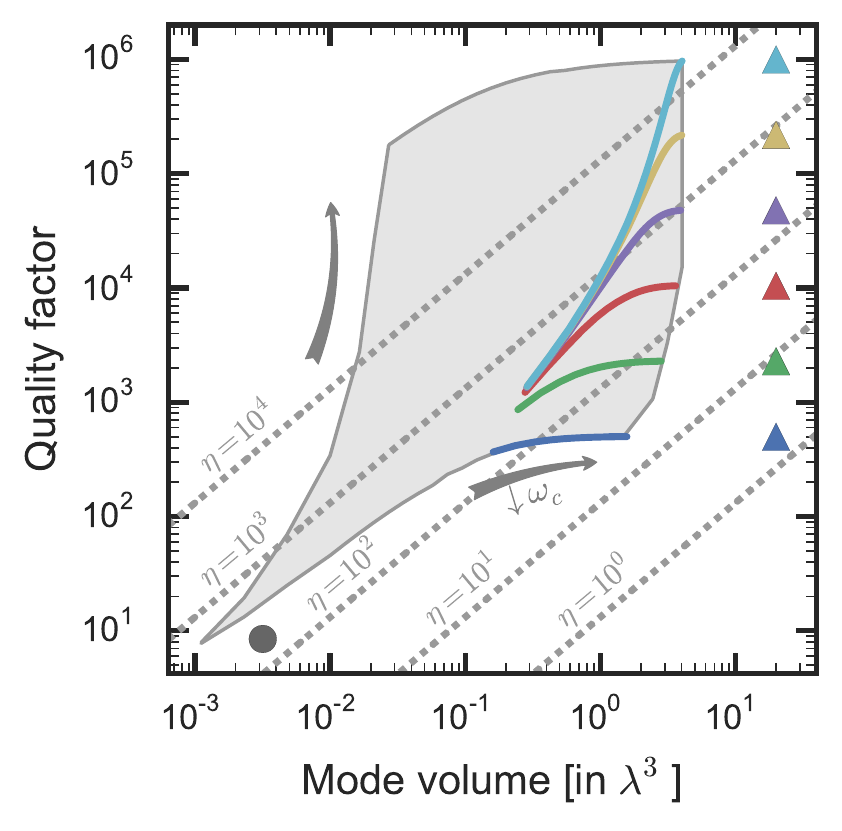} 
\caption{Phase diagram of quality factors $Q$ and dimensionless mode volumes $V/\lambda^3$. Shown are the values for the bare antenna (dark circle) and a set of bare cavities ($\blacktriangle$), as well as the values of the corresponding hybrid modes. The colored lines show hybrid results for all cavity-antenna detunings used. For decreasing $\ooc$, i.e. further red-detuning of the cavity, hybrid $Q$ and $V$ lie closer to those of the bare cavity. The light grey area indicates the location of the hybrid values attained for cavities with $500<Q<10^6$ and $0.5^3< \Veff / \lambda^3 <20$. Dashed grey lines are lines of constant emission enhancement $\eta$.} \label{fig:PhaseDiagram} 
\end{figure} 

Hybrid systems do not only offer increased emission enhancement compared to their bare constituents, they also open up an entirely new range of quality factors and mode volumes. \cref{fig:PhaseDiagram} show a `phase diagram' of $Q$ and $V$. Typically, plasmonic antennas are found in the bottom left of this diagram, with low $Q$ and low $V$. Cavities are usually found in the top right, with high $Q$ and high $V$. However, for most applications, being in either one of these extrema is not optimal. For example, if one desires a high Purcell factor, yet wants to avoid strong coupling --- demands that are critical to a good, low-jitter single photon source \cite{Lodahl2015} --- the high quality factors of cavities are unpractical, whereas antenna $Q$'s are so low that they may unnecessarily limit Purcell factors. A device with an intermediate $Q$ would be ideal, provided that the Purcell factor remains high. Another reason to want an intermediate $Q$, could be to match the bandwidth of enhancement to the emission spectrum of an emitter, which is often broader than that of a high-$Q$ cavity yet narrower than that of an antenna \cite{Englund2005}. Moreover, to obtain a device with an optimal trade-off between stability and tunability, one should be able to reach this regime of intermediate $Q$: high $Q$ renders cavities easily detuned by undesired perturbations, whereas the very low $Q$ of antennas makes them difficult to tune. Here we will show that hybrid systems allow precisely this: choosing the $Q$-factor to a desired, intermediate value, while retaining or even improving on the bare cavity Purcell factor. We generalize the previous results to cavities with a wide range of $Q$ and $\Veff$ to show the full attainable range of hybrid parameters. 

In \cref{fig:PhaseDiagram}, we compare $Q$ and $\Veff$ of modes in hybrid systems with those in the bare cavities and antenna. We assume the same antenna as in \cref{fig:EnhUL,fig:BreakdownTerms}. Cavities were used with $500<Q<10^6$ and $0.5^3<\Veff / \lambda^3<20$, and for each combination of $Q$ and $\Veff / \lambda^3 $ we take several different cavity resonance frequencies $100$ THz $<\ooc < 433$ THz, corresponding to cavity-antenna detunings ranging from 0.5 to 6.6 antenna linewidths. Cavities were always red-detuned from the antenna. The cut-off at 100 THz was chosen so as to stay within the realm of optical frequencies. To position hybrid structures in this diagram, we calculate emission enhancement for frequencies around the cavity resonance. We retrieve $Q$ from the linewidth of the Fano-resonance (see \cref{sec:SI-AlphaChi}). While mode volume is only well defined for a single (non-leaky) mode \cite{Koenderink2010,Kristensen2012,Sauvan2013}, here we employ an operational definition through Purcell's formula (\cref{Eq:Purcell}) and the peak value of the emission enhancement ($\ettot^{\mathrm{peak}}$). This leads to 
$\Veffhyb= \left( 3/(4\pi^2) \right) $Q$ / \ettot^{\mathrm{peak}} $,
with $\Veffhyb$ in units of the cubic resonance wavelength. We use the same definition for the antenna mode volume \footnote{In this section, when we speak about mode volumes, we always refer to the \emph{effective} mode volume.}. Note that, because we keep cavity $Q$ and $\Veff / \lambda^3$ constant when varying $\ooc$, cavities with different $\ooc$ appear as single points in \cref{fig:PhaseDiagram}. Hybrid $Q$ and $V$ however depend strongly on cavity-antenna detuning, as we have seen in \cref{sec:BareVsHyb}. Therefore the hybrid systems composed of cavities with different $\ooc$ appear as lines in \cref{fig:PhaseDiagram}.

From \cref{fig:PhaseDiagram} we see that hybrid systems provide exactly the tunability discussed earlier: it allows to choose any practical $Q$ between that of the cavity and the antenna, while keeping emission enhancement constant, or even improving it. The subset displayed in color shows that, compared to the bare cavity, one almost always plainly gains in terms of enhancement by adding the antenna. If the bare cavity provides an enhancement far below that of the antenna (blue and green), this gain can be very large, yet the $Q$-factor can be tuned only moderately. For cavities with enhancements similar to the bare antenna (red, purple and yellow), one can gain with respect to both bare components, and $Q$ can be tuned over a large range while maintaining very high enhancement. As can be expected, the Purcell factor of the cavities with highest $Q$ (light blue) is reduced by inclusion of the antenna, as cavities with such narrow resonances are easily spoiled by the losses introduced by an antenna. Yet it is remarkable that enhancements of order $10^3$ can be maintained over a large range of strongly reduced $Q$-factors in such systems. For all hybrid systems, the cavity-antenna detuning shifts the hybrid mode from a higher-$Q$ mode (for large detuning) to a lower-$Q$ mode (for small detuning). To illustrate the full attainable range of hybrid $Q$ and $V$, the light grey area shows where all the hybrid systems are located, for the full range of cavities examined here. From this we see that any $Q$ between that of the cavity and the antenna can be obtained.
Thus, hybrid systems can bridge the gap in $Q$ and $\Veff$ between cavities and plasmonic antennas, reaching any desired, practical $Q$ at similar or better enhancement factors.


\section{\label{sec:COMSOL}Finite-element simulations on a realistic hybrid system}

\begin{figure}[b]
\includegraphics[width=\figwidth]{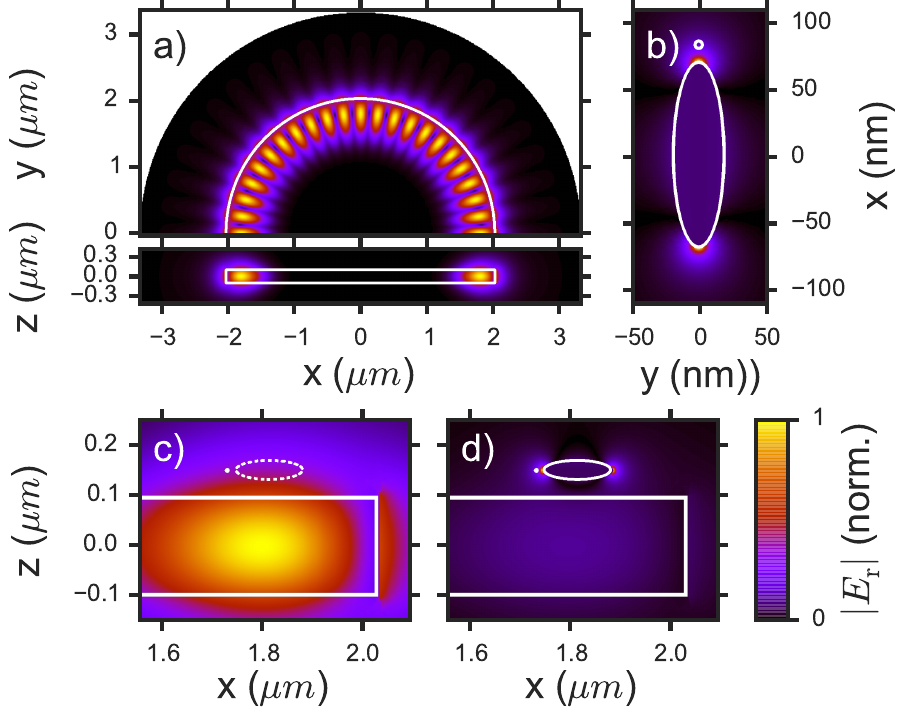} 
\caption{Cross-cuts of the cavity, antenna and hybrid mode profiles. All fields are normalized to their maximum values. Cross-cuts are taken at symmetry planes of the structures. White lines indicate the edges of the structures.
\textbf{a)} Top view and side view of the bare cavity eigenmode. Only the dominant (radial) field component is shown.
\textbf{b)} Field profile of the bare antenna in vacuum, illuminated by an x-polarized plane wave at its resonance frequency. The x-component of the scattered field is shown. The small white circle above the antenna tip indicates where we will place the source dipole. 
\textbf{c)} Zoom-in of the bare cavity eigenmode profile. The position of the antenna in the hybrid system is indicated with the dashed line. Note that no antenna was used in this simulation however. The position of the drive dipole is indicated beside the antenna tip.
\textbf{d)} Zoom-in of the hybrid eigenmode profile. Hot-spots are visible near the antenna tips.
} \label{fig:ModeProfiles} 
\end{figure}

Here we demonstrate a possible physical implementation of the proposed hybrid systems. We perform finite-element simulations on a realistic cavity-antenna design using COMSOL Multiphysics, version 5.1. These simulations also serve to verify the validity of our analytical oscillator model. \cref{fig:Cartoon} is an artistic representation of the simulations. 
As a cavity, we take a silicon nitride (n=1.997) disk in vacuum with a radius of 2032 nm and a thickness of 200 nm. A small amount of absorption was added by introducing an imaginary component (4$\cdot$ 10$^{-6}$) to the permittivity of the silicon nitride. This damping allows to tune the cavity Q, and helps to trace in the simulation how much power flows into the cavity mode. The disk supports a radially polarized $m$=22 whispering gallery mode (WGM) at 382.584 THz ($\sim$784 nm) with $Q$=7.28$\cdot$ 10$^4$ (see \cref{fig:ModeProfiles}a and c). The antenna we use is a gold prolate ellipsoid with a long (short) axis radius of 70 (20) nm. Optical constants are described by a modified Drude model \cite{Penninkhof2008}. \cref{fig:ModeProfiles}b shows the antenna field profile at resonance. 
    
To verify the predictions of the oscillator model, we first calculate emission enhancement spectra for the bare components, and through a fit retrieve from these all the necessary input parameters for our oscillator model. We then use the oscillator model to make a prediction for the enhancement spectrum of the hybrid system, and compare this to the enhancement spectrum obtain from a finite-element simulation of the hybrid system. 
To fit the bare component spectra we require expressions that derive from the equations of motion (\cref{Eq:EOMone,Eq:EOMtwo}), yet can be directly compared to the the power radiated and dissipated by the antenna, as well as the power outflux through the cavity loss channels, i.e. radiation and dissipation in the nitride. The latter two are characterized by the radiative and absorptive damping rates $\kr$ and $\kabs$, respectively, with $\kk=\kr+\kabs$. We derive expressions for the antenna radiation and absorption using the equations of motion \cref{Eq:EOMone,Eq:EOMtwo}, by calculating the work done by the Abraham-Lorentz force and the dissipative force, respectively, on the antenna (see \cref{sec:SI-SplitByChannel}). Expressions for the cavity losses are derived by multiplying the respective loss rates with the energy in the cavity mode, which we can obtain from the equations of motion. Importantly, these expressions not only let us fit the simulated spectra, but they also allow to study which channels the emitted power in hybrid systems flows into. This will be the topic of the next section. 

From the fit to the bare cavity emission and absorption spectra (see \cref{sec:SI-COMSOL}), we find the cavity parameters $\ooc$, $\kr/2\pi=5$ GHz, $\kabs/2\pi=0.3$ GHz and $\Veff = 22.8 \lambda^3$. This leads to a peak enhancement of 242. The bare antenna spectra yield the antenna parameters $\oz/2\pi=436$ THz, $\gi/2\pi= 18.1$ THz, $\beta=$0.073 C$^2$/kg and an effective source-antenna distance of 55.2 nm. The latter is the distance that enters the Green's function in \cref{Eq:ettot}, which can be smaller than the actual distance (the dipole was placed 12 nm from the tip of the antenna, as shown in \cref{fig:ModeProfiles}b) due to the lightning rod effect: the sharp tips create larger field enhancements than expected from a normal dipole field. These values lead to a bare radiative (absorptive) antenna emission enhancement of 186 (174) at maximum.

\begin{figure}[b]
\includegraphics[width=\figwidth]{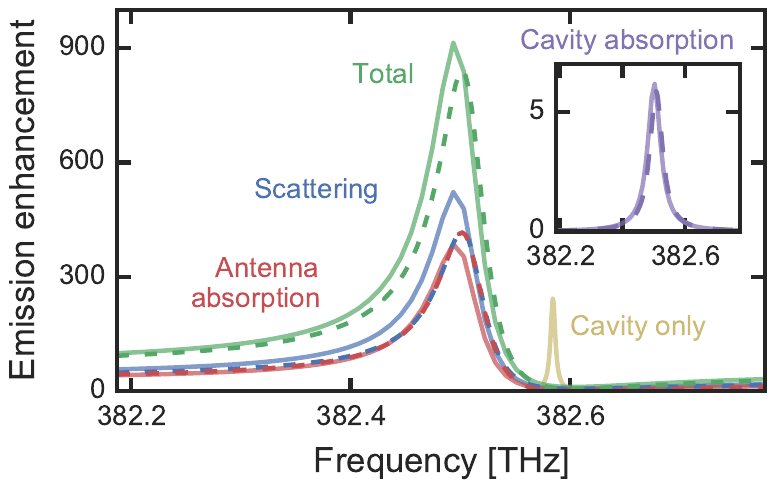} 
\caption{Emission enhancements in a hybrid system from the oscillator model (dashed) and from simulations (solid). We show enhancements due to scattering into free space (blue), antenna absorption (red) and total enhancement (green). Enhancement due to cavity absorption (purple) in the hybrid system is visible in the inset. Enhancement from the bare cavity (yellow) is shown for comparison.} \label{fig:COMSOLComparison} 
\end{figure} 

With the cavity and antenna parameters known, we can use the oscillator model to predict the emission enhancement in the hybrid system. \cref{fig:COMSOLComparison} shows the comparison between this prediction and the results from simulations on the hybrid system. For these, the antenna was placed beside the source, just above the disk, as shown in \cref{fig:ModeProfiles}d. We find an emission enhancement of $\sim$914 in the hybrid system, which is a large increase with respect to the bare cavity (242) and antenna (360 at resonance and $\sim$65 near cavity resonance). Moreover, the bandwidth over which this enhancement occurs is increased by a factor 9.4 (to 49 GHz) with respect to the cavity. There is excellent agreement between the model and the simulation, not only in the total enhancement but also in the separate radiative and absorptive components, as well as the cavity absorption. Remaining differences can be largely attributed to errors in the antenna fit (see \cref{fig:Fits}). The overlap between predicted radiative and absorptive enhancements is coincidental and a result of the imperfect antenna fit.
These results demonstrate that the oscillator model correctly predicts emission enhancement in a coupled cavity-antenna system, based on the response of the bare components. Moreover, it shows that also realistic cavity-antenna systems can combine the best features of both cavity and antenna, achieving much stronger emission enhancement than the bare components.

\section{\label{sec:Efficiency}Efficiency of radiation into the cavity}

\begin{figure}[t]
\includegraphics[width=\figwidth]{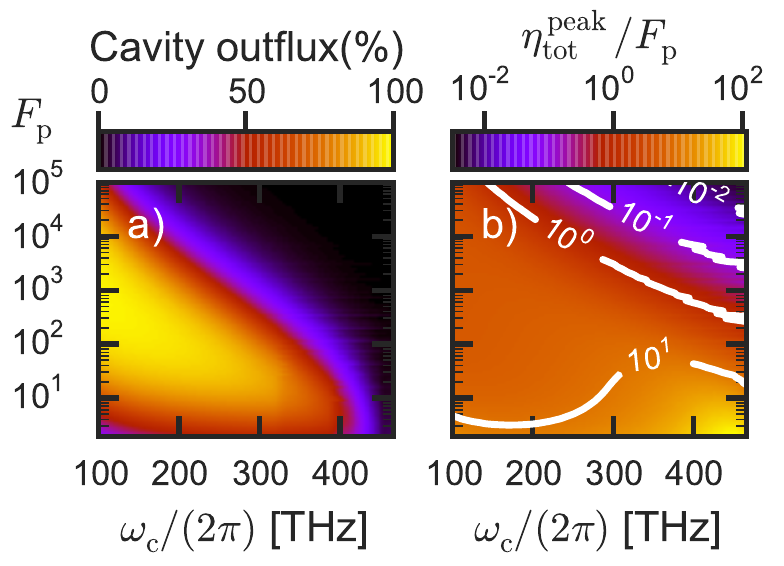} 
\caption{
\textbf{a)} Fraction of power into the cavity decay channel $\kk$, as function of cavity resonance $\ooc$ and bare cavity Purcell factor $\Fp$. This fraction was evaluated at the peak of the total emission enhancement $\ettot$. We use the same antenna as in \cref{fig:EnhUL,fig:PhaseDiagram}. 
\textbf{b)} Total emission enhancement $\ettot^{\mathrm{peak}}$ of the hybrid system (at peak), relative to $\Fp$. The same cavities and antenna were used as in a).
} \label{fig:ComboMap} 
\end{figure} 

In the previous sections, we demonstrated that hybrid systems allow strongly boosted emission enhancements at any desired quality factor $Q$. Here we will show that, by hybridization, one can control into what channels energy is emitted. Depending on the application, one may e.g. wish to design a system that emits all power into free-space, or into a single-mode waveguide. The latter is often the case for an on-chip single-photon source, for example. In the previous section we established that we are able to separate the powers going into the different decay channels of the system, i.e. antenna radiation or absorption and the loss channels intrinsic to the cavity. Here we use the expressions for these powers to study the fraction of power going into the cavity decay channel, as this is usually most efficiently extracted in e.g. a waveguide. This fraction, i.e. the efficiency of extraction into single mode output channel, is also known as the $\beta$-factor in the context of single-photon sources \cite{Lodahl2015}. Note that, as we generally have not specified the origin of the cavity loss $\kk$, one could assume it to be dominated by outcoupling to a waveguide. In experiments this is commonly achieved by evanescent coupling of a cavity to a nearby integrated waveguide or fiber taper \cite{Knight1997,Spillane2003}. Overcoupling then ensures that the waveguide or taper is the dominant loss channel.

\cref{fig:ComboMap}a and b show the relative cavity outflux and the peak value of the total hybrid emission enhancement $\ettot^{\mathrm{peak}}$ as function of cavity resonance $\ooc$ and bare cavity Purcell factor $\Fp$. The same cavities and antenna were used as in \cref{fig:PhaseDiagram}, and detuning now ranged between 0 and 6.6 antenna linewidths. Note that relative cavity outflux and $\ettot^{\mathrm{peak}}$ are fully determined by antenna properties, detuning and $\Fp$ (i.e. $Q$/$\Veff$), not by $Q$ and $\Veff$ separately. There is a large region in which hybrid emission enhancement can be increased with respect to the cavity, while maintaining a very high fraction of power flux into the cavity channel. This implies that the plasmonic antenna helps to boost emission enhancement though its boosted field enhancement, while adding almost no additional losses. It is consistent with the results in \cref{fig:EnhUL}c, where broadening by the antenna quickly decays towards low frequencies, whereas confinement remains significant. \cref{fig:ComboMap} shows that this works particularly well for cavities with $\Fp$ between 10 and $\sim$10$^3$. Close to the antenna resonance (460 THz), cavity outflux drops, as power outflux is dominated by the antenna. For very good cavities with $\Fp$ around 10$^4$, power outflux is also dominated by the antenna, even for considerably far red-detunings. This reflects the fact that either intrinsic cavity losses are very low (high $Q$), or coupling to the antenna is very strong (low $V$). Both cases lead to the antenna decay channels being dominant. Importantly, dominant outcoupling through the antenna does not mean that all the power is dissipated: it is distributed between dipolar radiation and dissipation according to the bare antenna albedo. For applications where radiative efficiency rather than coupling to a waveguide is important, for example, these antenna-dominated regimes can  be highly interesting. In conclusion, one can generally engineer the system in such a way that the power flows in any of the desired channels. Specifically, we have shown that it can be designed for a high extraction efficiency into a single cavity loss channel, such as a waveguide. This is of particular interest for applications such as an on-chip single-photon source with a high $\beta$-factor.

\section{Conclusion}
We have shown that hybrid cavity-antenna systems can achieve larger emission enhancement than either the antenna or the cavity alone. These systems can benefit simultaneously from the high cavity quality factor and the low mode volume of the antenna. Surprisingly, this benefit occurs only when the cavity is red-detuned from the antenna. We have demonstrated that this is partly due to the frequency-dependent radiation damping of the antenna, and partly because at these detunings, radiation by the emitter into the cavity constructively interferes with scattering by the antenna. Moreover, we have compared quality factors and mode volumes of resonances in hybrid systems to those in the bare antenna and cavity, showing that hybrid structures allow to design any desired quality factor while maintaining similar or higher emission enhancement than the bare components. A study of the cavity power outflux as a fraction of total emitted power demonstrated that one can furthermore engineer the system to emit efficiently into a desired output channel. It was shown that there is a large range of cavities that allow simultaneously boosting the emission enhancement and maintaining a high extraction efficiency into a single cavity loss channel, allowing efficient outcoupling into a waveguide. Finally, a physical implementation using a WGM cavity and a gold antenna was proposed and tested using finite-element simulations, showing strongly increased emission enhancement and excellent agreement with the oscillator model.

\begin{acknowledgments}
This work is part of the research program of the Foundation for Fundamental Research on Matter (FOM), which is financially supported by the Netherlands Organization for Scientific
Research (NWO). The authors gratefully acknowledge Ben van Linden van den Heuvell, Klaasjan van Druten, Robert Spreeuw, Francesco Monticone and Andrea Al\`{u} for inspiring discussions, and Henk-Jan Boluijt for the designs used in \cref{fig:Cartoon}.
\end{acknowledgments}

\appendix
\crefalias{section}{appsec}

\section{\label{sec:SI-EOMS}Equations of motion for a cavity-antenna system}
Here we derive the equations of motion (\cref{Eq:EOMone,Eq:EOMtwo}) for a coupled cavity-antenna system. We model the antenna as a point dipole with the familiar Lorentzian polarizability, as found e.g. for the Fr\"{o}hlich mode of a small metal sphere \cite{Bohren} in vacuum. Radiation damping is included to make the model self-consistent and adhering to the optical theorem \cite{DeVries1998}. Interaction with the cavity mode is explicitly separated out from this radiation damping due to other modes, and included in a second equation of motion. This equation of motion, describing a single cavity mode, is based on temporal coupled mode theory. Its derivation is analogous to deriving the classical equation of motion for an atom in a cavity, as given by Haroche \cite{Haroche1992}, where in our work the 'atom' will be representing the antenna. Throughout this derivation, all quantities are in SI units.

\subsection{A dipolar antenna}
We consider a system of a small nanoantenna positioned in the field of a cavity at position $\ro$. The antenna is described as a point dipole with dipole moment $\p = p \hatp$, where $\hatp$ is the unit vector pointing along $\p$, and we assume for simplicity that it is only polarizable along $\hatp$. This analysis can be easily extended to a tensor polarizability, however. 

The antenna response is modeled as a harmonic oscillator of charge $q$ and mass $m$ with resonance frequency $\oo_0$ that suffers from intrinsic damping due to Ohmic heating described by an energy damping rate $\gi$. The equation of motion (EOM) that governs the time dependence of the (complex) dipole amplitude $p(t)$ is that of a damped, driven harmonic oscillator:
\begin{equation}
\ddot{p} + \gi\dot{p} + \oo_0^2 p = \beta E,\label{Eq:EOMp1}
\end{equation}
where $\beta = q^2/m$ is the oscillator strength and $E = \E(\ro)\cdot\hatp$ is the total electric field $\E$ present at the antenna position, projected on $\hatp$. We will separate it in three contributions:
\begin{equation}
E = E_c + E_p + \Epdr.\label{Eq:Eseparation}
\end{equation}
Here, $E_c$ is the field of the cavity mode of interest at $\ro$, along the dipole direction. The second term is the field at the antenna, caused by the antenna itself. It can be formally written as
\begin{equation}
E_p(t) = \int_{-\infty}^{\infty}\!\mathrm{d}t'\,\mathcal{G}_\mathrm{bg}(t-t')p(t') = \left(\mathcal{G}_\mathrm{bg}\ast p\right)(t),
\end{equation}
where $\mathcal{G}_\mathrm{bg}\!\left(t-t'\right)$ is a linear response function that describes the field at the position of the antenna at time $t$ due to a delta function excitation at time $t'$. 
Its Fourier transform is $\Gbg\!\left(\ro,\ro,\oo\right)\equiv\hatp\,\cdot\! \Gr_\mathrm{bg}\!\!\left(\ro,\ro,\oo\right)\!\cdot\!\hatp$; the projection along the antenna direction of the Green's tensor $\Gr_\mathrm{bg}$ that describes the field $\E_p$ of the antenna in its environment via 
$ \E_p\!\left(\r,\oo\right) = \Gr_\mathrm{bg}\!\!\left(\r,\ro,\oo\right)\cdot\mathbf{p}\left(\oo\right). \label{Greendefinition} $
Importantly, we need to explicitly omit the contribution of the cavity field in this response, since that will be accounted for in the next section. Instead, it is composed of the antenna radiation expanded in all modes \emph{except} the cavity mode. It is for that reason that we use the subscript `bg' to mean that only the dielectric `background' contributes to $\mathcal{G}_\mathrm{bg}$. This dielectric background can in principle be inhomogeneous, and as such the response can be altered from that in a homogeneous medium due to the excitation of for example modes in a substrate or other cavity modes. If those contributions are negligible, the well-known expression for the Abraham-Lorentz force in a homogeneous medium can be used such that
\begin{equation}
E_p = \frac{\sqrt{\ee}\, \dddot{p}}{6\pi\eo c^3},
\end{equation}
where $\ee=\ee(\ro)$ is the relative permittivity of the medium \cite{Novotny2012}.

The final term $\Epdr$ in \cref{Eq:Eseparation} is the external driving field, i.e. the electric field at the position of the antenna which does \emph{not} find its origin in the antenna itself, and is distributed over other modes than the cavity mode. This can for example be the field due to an oscillating source dipole.

\subsection{A cavity}
Next, we seek to find a similar expression for the cavity field $\Ec$. First, we must assume that the field can be expanded in orthogonal modes $\mathbf{E}_m$, of which the cavity mode $\mathbf{E}_\mathrm{c}$ is just one. This is a standard approach to describe the physics of high-Q cavities in quantum optics. We note that for very open systems, there is currently a strong debate about quasi-normal modes appropriate for non-hermitian systems \cite{Kristensen2012,Sauvan2013,Yang2015}. We note that generalization of our formalism to deal with quasi-normal modes is outside the scope of this work. Such a generalization would also require to revisit  the definitions of mode normalization, inner product, and energy density.
The assumed orthogonal eigenmodes can be factorized as
\begin{equation}
\Em\!\left(\r,t\right) = a_m(t)\mathbf{e}_m(\r), \label{separatingvariables}
\end{equation}
normalized such that
\begin{equation}
\int\!\mathrm{d}\r\,\frac{1}{2} \eo\ee(\r) \mathbf{e}_m^\ast(\r)\cdot\mathbf{e}_n(\r)=\delta_{mn} \label{orthonormality}
\end{equation}
and
\begin{equation}
\left|a_m\right|^2 = \int\!\mathrm{d}\r\,\frac{1}{2} \eo\ee(\r) \left|\Em\!\left(\r,t\right)\right|^2 = U_m , \label{Eq:SI-Um}
\end{equation}
with $U_m$ the total energy of the mode. The EOM for the time-dependent part of the cavity mode $\ac(t)$ can then be derived as \cite{Haroche1992}
\begin{equation}
\ddot{\ac}+\kk \dot{\ac} + \ooc^2 \ac = -\frac{1}{2}\hatp\cdot\mathbf{e}_\mathrm{c}^\ast(\ro)\ddot{p}, \label{Eq:EOMac}
\end{equation}
with $\kk$ the damping rate of the cavity mode $\mathbf{E}_\mathrm{c}$ and $\ooc$ its eigenfrequency. Multiplying \cref{Eq:EOMac} with $\hatp\cdot\mathbf{e}_\mathrm{c}(\ro)$ and introducing the effective mode volume
\begin{equation}
\Veff = \frac{\int\!\mathrm{d}\r\, \ee(\r)\left|\mathbf{E}_\mathrm{c}(\r)\right|^2}{\ee(\ro)\left|\hatp\cdot\mathbf{E}_\mathrm{c}(\ro)\right|^2} 
= \frac{2}{\eo\ee(\ro)\left|\hatp\cdot\mathbf{e}_\mathrm{c}(\ro)\right|^2}, \label{Eq:SI-Veff}
\end{equation}
we obtain
\begin{equation}
\ddot{\Ec}+\kk \dot{\Ec} + \ooc^2 \Ec + \frac{1}{\eo \ee \Veff } \ddot{p} = 0 , \label{Eq:EOMEc1}
\end{equation}
where $\Ec = \hatp\cdot\mathbf{e}_\mathrm{c}(\ro) \ac $ is the cavity mode field projected on the antenna axis, and $\ee=\ee(\ro)$. Note that $\Veff$ is the effective mode volume as it is felt by the antenna at position $\ro$, and it is therefore tunable by moving the dipole in the cavity mode. In that respect it differs from the more usual definition of a cavity mode volume that uses the \emph{maximum} field in the cavity mode instead.

\subsection{Equations of motion in the Fourier domain}
\cref{Eq:EOMp1,Eq:EOMEc1} are most easily solved in the Fourier domain, where they result in
\begin{align}
\left(\oz^2-\oo^2-\ii\oo\gi-\beta\Gbg\!\left(\ro,\ro,\oo\right) \right) p - \beta \Ec &= \beta \Epdr, \\
- \frac{\oo^2}{\eo \ee \Veff } p + \left(\ooc^2-\oo^2-\ii\oo\kk\right) \Ec &= \oo^2 \Ecdr.
\end{align}
We have included a driving term $\oo^2 \Ecdr$ in the latter, which allows for an arbitrary field driving the cavity mode. Absorbing the real part of $\Gbg\!\left(\ro,\ro,\oo\right)$ in $\oz$ and the imaginary part in the total antenna damping rate $\gamma$, such that \cite{DeVries1998}
\begin{equation}
\gamma=\gi+\gr = \gi + \frac{\beta}{\oo}\Im{\Gbg\!\left(\ro,\ro,\oo\right)}
\end{equation}
with $\gr$ denoting the radiative damping rate, we retrieve the equations of motion as given  by \cref{Eq:EOMone,Eq:EOMtwo}, i.e.
\begin{align}
\left(\oz^2-\oo^2-\ii\oo\gamma \right) p - \beta \Ec &= \beta \Epdr, \label{Eq:fourierEOMp} \\
- \frac{\oo^2}{\eo \ee \Veff } p + \left(\ooc^2-\oo^2-\ii\oo\kk\right) \Ec &= \oo^2 \Ecdr, \label{Eq:fourierEOMEc}
\end{align}
Note that, using the relation between $\Im{\Gbg\!\left(\ro,\ro,\oo\right) }$ and the partial local density of states (LDOS) of the background $\rbg$, the radiative damping rate $\gr$ may also be expressed as \cite{Novotny2012}
\begin{equation}
\gr = \frac{\beta \pi }{6 \eo \ee} \rbg. \label{Eq:gr}
\end{equation}
In our calculations, we use the vacuum LDOS $\rho_\mathrm{vac} = \oo^2 / ( \pi^2 c^3) $ for $\rbg$.

\section{\label{sec:SI-ettot}Total emission enhancement}
Here we derive expressions for the total emission enhancements of a hybrid system (\cref{Eq:ettot}) and of the bare antenna and bare cavity. The latter are used for the bare component spectra in \cref{fig:EnhUL}a. 

\subsection{\label{sec:SI-TotalEnh-DipoleDrive}Driving by a dipolar source}
To study how an emitter would behave in the hybrid system, we now continue to identify $\Epdr$ and $\Ecdr$ with the driving fields from a dipolar source. We consider a small, non-polarizable, dipole with harmonically oscillating dipole moment $\pdr \hatp_\mathrm{dr}$ at position $\rdr$ as a `constant current' driving source. This source dipole drives the antenna with a field
\begin{equation}
\Epdr=\Gbg\!\left(\ro,\rdr,\oo\right) \pdr = \Gbg \pdr ,
\end{equation}
where $\Gbg\!\left(\ro,\rdr,\oo\right) = \hatp\,\cdot\!\Gr_\mathrm{bg}\!\!\left(\ro,\rdr,\oo\right)\cdot\hatp_\mathrm{dr}$.
The emitter being a point source, it will be able to drive all modes that have non-zero electric field at its position, including the cavity mode. If we redo the derivation of the cavity equation of motion, including a term that accounts for excitation by the dipole, we find
\begin{equation}
- \frac{\oo^2}{\eo \ee \Veff } p + \left(\ooc^2-\oo^2-\ii\oo\kk\right) \Ec = \frac{\oo^2}{\eo \ee \Veff} \phi \, \pdr,
\end{equation}
where $\phi= \left( \hatp_\mathrm{dr} \cdot \mathbf{e}_\mathrm{c}^\ast(\rdr) \right) / \left( \hatp \cdot \mathbf{e}_\mathrm{c}^\ast(\ro) \right)$ is a complex factor accounting for a difference in orientation between $\hatp$ and $\hatp_\mathrm{dr}$, as well as a different cavity mode field at $\ro$ and $\rdr$. Comparison with \cref{Eq:fourierEOMEc} leads to
\begin{equation}
\Ecdr=\frac{\phi}{\eo \ee \Veff } \pdr.
\end{equation}
If we take the source to be polarized along the antenna axis, such that $\hatp\!\cdot\!\hatp_\mathrm{dr}=1$, and we assume that the cavity field is equal at $\ro$ and $\rdr$ (e.g. because source and antenna are very close compared to the wavelength), we obtain
\begin{equation}
\Ecdr=\frac{1}{\eo \ee \Veff } \pdr.
\end{equation}
The EOMs including the dipolar driving terms become
\begin{align}
\left(\oz^2-\oo^2-\ii\oo\gamma \right) p - \beta \Ec &= 
\beta \Gbg \pdr , \label{Eq:fourierEOMpDip} \\
- \frac{\oo^2}{\eo \ee \Veff } p + \left(\ooc^2-\oo^2-\ii\oo\kk\right) \Ec &= 
\frac{\oo^2}{\eo \ee \Veff } \pdr. \label{Eq:fourierEOMEcDip}
\end{align}

\subsection{Total emission enhancement}
To calculate the emission enhancement for the drive dipole, we must know the field returning at the dipole position after interaction with the system. Let us first use \cref{Eq:fourierEOMpDip,Eq:fourierEOMEcDip} to express the antenna dipole moment $p$ in terms of the drive dipole amplitude $\pdr$ by eliminating the cavity field $\Ec$ from the equations. We obtain
\begin{align}
p &= \frac{\beta}{\oz^2 -\oo^2 - \ii\oo \gamma - \beta \chihom} \left( \Gbg + \chihom \right) \pdr \\
&= \ahyb \, \left( \Gbg + \chihom \right) \pdr , \label{Eq:phyb}
\end{align}
where $\ahyb$ is the hybridized antenna polarizability given by \cref{Eq:ahyb}. We see that $p$ is polarized in response to both the direct excitation by the source ($\Gbg \pdr$) and the cavity field ($\chihom \pdr$). However, it responds with a hybridized polarizability, due to coupling with the cavity.
With $p$ known, we can then express the field scattered by the antenna at the position of the source dipole as:
\begin{equation}
E_{\mathrm{s}}(\rdr) =  \Gbg (\rz , \rdr , \omega) \, p = \Gbg \, p,
\end{equation}
where we have used reciprocity, i.e. $\Gbg (\rz , \rdr , \omega) = \Gbg (\rdr , \rz , \omega)$. 

Similarly, we can eliminate $p$ from \cref{Eq:fourierEOMpDip,Eq:fourierEOMEcDip} to express $\Ec$ as a function of $\pdr$:
\begin{align}
\Ec &= \frac{\oo^2}{\eo \ee \Veff} \frac{( 1+ \ahom \Gbg ) \pdr}{\ooc^{2}-\omega^{2}-\ii \omega \gamma - \dfrac{\oo^2}{\eo \ee \Veff} \ahom} \\
 &= \chihyb \, ( 1+ \ahom \Gbg )\pdr, \label{Eq:Echyb}
\end{align}
with $\chihyb$ being the hybridized cavity response given by \cref{Eq:chihyb}. Similar to the situation in \cref{Eq:phyb}, we recognize that the cavity is excited by both the source and the induced dipole moment of the antenna, and responds with the hybridized cavity response $\chihyb$. The cavity field returning at the source position is equal to $\Ec$.
We can now express the total field at the location of the source as:
\begin{equation}
E_{\mathrm{tot}}(\rdr) = E_{\mathrm{hom}}(\rdr) + E_{\mathrm{s}}(\rdr) + \Ec,
\end{equation}
where $E_{\mathrm{hom}}(\rdr) = \Gbg\!(\rdr , \rdr , \omega) \pdr $ is the field that has interaction with the homogeneous background medium only (i.e. the field responsible for the well-known expression for dipole radiation in a homogeneous medium). 

\begin{figure}[!hbt]
\begin{center}
\includegraphics[width=\figwidth]{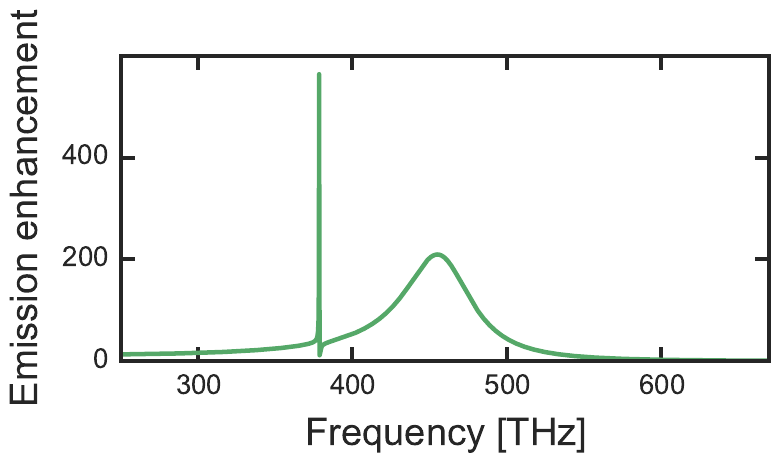} 
\caption{Emission enhancement spectrum for a hybrid system. We have taken the same antenna and a cavity with the same quality factor and dimensionless mode volume as used for \cref{fig:EnhUL}. In contrast to \cref{fig:EnhUL}, where curves for multiple systems with different antenna-cavity detunings were plotted, we show only one system with a cavity red-detuned from the antenna by 1.5 antenna linewidths (81.4 THz). The spectrum contains a narrow peak and a broad peak, corresponding to the 'cavity-like' and the 'antenna-like' eigenmode of the hybrid system, respectively.} \label{fig:SI-Hybridspectrum} 
\end{center}
\end{figure}

The cycle-averaged work done by the driving source can be found by calculating the work done by $E_{\mathrm{tot}}(\rdr)$ on the source itself \cite{Novotny2012}:
\begin{align}
\Pdr &= \frac{\omega}{2} \Im{\pdr^{*} \cdot E_{\mathrm{tot}} }\\
&= \frac{\omega }{2} \, |\pdr|^2 \, \mathrm{Im}  \{ \, \Gbg\!(\rdr , \rdr , \omega) + \ahyb \Gbg^2  \notag \\
&\qquad \qquad + \Gbg \ahyb \chihom +\chihyb \ahom \Gbg +\chihyb \} \label{Eq:Pdr1}.
\end{align}
It can be straightforwardly shown that $\ahyb \chihom = \ahom \chihyb$, such that this expression further simplifies to:
\begin{align}
\Pdr = \frac{\omega }{2} \, |\pdr|^2 \, \mathrm{Im}  \{ \, \Gbg (\rdr , \rdr , \omega) + \ahyb \Gbg^2  \notag \\ 
+ 2 \Gbg \ahyb \chihom +\chihyb \}. \label{Eq:Pdr2}
\end{align}
To arrive at emission enhancement, one should calculate the ratio of this power and that emitted by the same dipolar source in a homogeneous medium. The latter is given by Larmor's formula and is equal to the contribution of the first term in \cref{Eq:Pdr2}:
\begin{equation}
P_{\mathrm{hom}}= \frac{\omega^4 n}{12 \pi \eo c^3 } |\pdr|^2, \label{Eq:Larmor}
\end{equation}
with $n$ the refractive index of the medium. The total emission enhancement experienced by the source dipole is thus
\begin{align}
\ettot &= \frac{\Pdr}{P_{\mathrm{hom}}} \notag \\
&= 1+ \frac{6 \pi \eo c^3 }{\oo^3 n } \Im{\ahyb \Gbg^2  + 2 \Gbg \ahyb \chihom +\chihyb }, \label{Eq:ettot_hyb}
\end{align}
which corresponds to \cref{Eq:ettot} in the main text. \cref{fig:SI-Hybridspectrum} shows an example of a hybrid enhancement spectrum calculated using \cref{Eq:ettot_hyb}. 

Finally, the emission enhancements for a dipolar source coupled to a \emph{bare} antenna or cavity can, by taking respectively $\Veff \rightarrow \infty$ or $\beta \rightarrow 0$, be straightforwardly found as
\begin{align}
\etptot &= 1+ \frac{6 \pi \eo c^3 }{\oo^3 n } \Im{\ahom \Gbg^2}. \label{Eq:etptot} \\
\etctot &= 1+ \frac{6 \pi \eo c^3 }{\oo^3 n } \Im{\chihom}. \label{Eq:etctot}
\end{align}

\section{\label{sec:SI-AlphaChi}The hybridized polarizability and cavity response}
In this section we discuss the shape of the hybridized antenna polarizability $\ahyb$ and the hybridized cavity response $\chihyb$ (\cref{Eq:ahyb,Eq:chihyb} in the main text). We show that $\ahyb$ takes on a Fano-type lineshape near the cavity resonance, which agrees with the results by Frimmer et al. \cite{Frimmer2012}. The cavity response $\chihyb$ has a Lorentzian lineshape that is shifted and broadened with respect to the bare cavity resonance in a manner consistent with Bethe-Schwinger perturbation theory \cite{Bethe1943,Waldron1960}.

\begin{figure}[!hbt]
\begin{center}
\includegraphics[width=\figwidth]{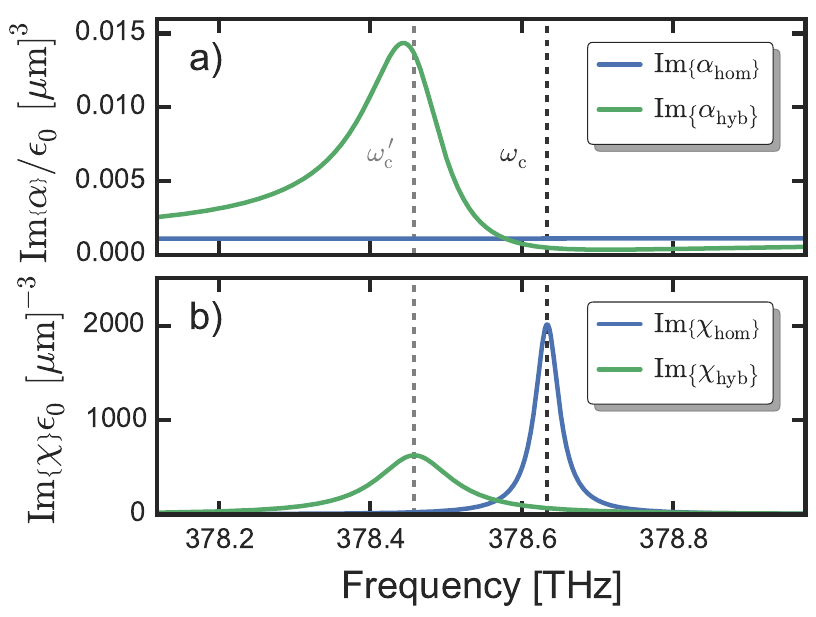} 
\caption{
\textbf{a)} Example of the bare antenna and hybridized polarizabilities $\ahom$ and $\ahyb$. We take the same antenna and cavity parameters as used in \cref{fig:EnhUL} of the main paper, and take the cavity red-detuned from the antenna by 1.5 antenna linewidths. While the bare polarizability is virtually constant, $\ahyb$ shows a Fano line shape. Dashed lines indicate the bare and hybridized cavity resonance frequencies $\ooc$ and $\ooc'$, respectively. 
\textbf{b)} Bare and hybridized cavity responses $\chihom$ and $\chihyb$, for the same system as used in a. Contrary to the hybridized polarizability, $\chihyb$ does not have a Fano lineshape, but rather that of a Lorentzian resonance, shifted and broadened compared to the bare cavity resonance.} \label{fig:AlphaChi} 
\end{center}
\end{figure}

Starting from the definition of $\ahyb$, i.e. \cref{Eq:ahyb} in the main text, we can take the limit of a low-loss cavity and evaluate close to the cavity resonance frequency $\ooc$ (i.e. $\Delta, \kk \ll \ooc$, with $\Delta \equiv \oo - \ooc$). This allows us to write $\ahyb$ in the familiar shape of a Fano resonance \cite{Novotny2012}, i.e.
\begin{equation}
\ahyb \approx \ahom(\ooc) \left( 1 + F \frac{ \kk' /2}{-\ii(\oo - \ooc') + \kk'/2} \right). \label{Eq:SI-ahybFano}
\end{equation}
Here, $F$ is the Fano-parameter
\begin{equation}
F = \frac{i \OO^2 \ooc }{\kk' \left( \oz^2 -\ooc^2 -\ii \ooc \gamma \right)} , 
\end{equation}
with $\OO=\sqrt[]{\beta / (\eo \ee \Veff)}$ the antenna-cavity coupling rate. \cref{Eq:SI-ahybFano} also shows the hybridized eigenfrequency $\ooc'$ and damping rate $\kk'$, given as 
$\ooc' = \ooc + \delta \ooc$ and $\kk' = \kk + \delta \kk$, with 
\begin{align}
\delta \ooc &= -\frac{\OO^2 \ooc \left( \oz^2 -\ooc^2 \right)}{2 \left( \left( \oz^2-\ooc^2 \right)^2 +\ooc^2 \gamma^2 \right) } \notag \\
&= - \frac{ \ooc }{2 \eo \ee \Veff} \Re{\ahom (\ooc)}, \label{Eq:doc}\\
\delta \kk &= \frac{\OO^2 \ooc^2 \gamma}{\left( \oz^2-\ooc^2 \right)^2 +\ooc^2 \gamma^2}  
= \frac{ \ooc }{\eo \ee \Veff} \Im{\ahom (\ooc)} \label{Eq:dkk}.
\end{align}
F determines the shape of the Fano-resonance: if it is real and positive, the resonance (i.e. $\Im{\ahyb}$) takes on a Lorentzian line shape. If it is real and negative, interference is entirely destructive. Anywhere in between gives an asymmetric line shape. An example of such an asymmetric resonance is shown in \cref{fig:AlphaChi}a.
The Fano shape stems from the interference between the two new eigenmodes of the hybrid system, of which one is broad and similar to the unperturbed antenna mode and the other is narrow and similar to the unperturbed cavity mode. The latter has eigenfrequency and -damping $\ooc'$ and $\kk'$, respectively.

With the same assumptions as above, we can rewrite the expression for $\chihyb$ (\cref{Eq:chihyb} in the main text) in the form:
\begin{equation}
\chihyb \approx \frac{1}{\eo \ee \Veff} \frac{\ooc' / 2}{-(\oo - \ooc') - \ii \kk'/2} . \label{Eq:chihyb-approx}
\end{equation} 
This does not represent a Fano line shape, yet instead describes a Lorentzian with resonance frequency $\ooc'$ and damping rate $\kk'$ (as shown in \cref{fig:AlphaChi}b). We can see from \cref{Eq:doc,Eq:dkk} that this resonance is shifted and broadened with respect to the unperturbed cavity response $\chihom$. In fact, the expressions for the lineshift and broadening exactly match those found by Bethe-Schwinger cavity perturbation theory. This is consistent with the fact that we have explicitly assumed in the derivation of the equations of motion that there is no far-field interference between the antenna and the cavity radiation. Under these assumptions Bethe-Schwinger perturbation theory holds \cite{Ruesink2015}.

\section{\label{sec:SI-SplitByChannel}Emission enhancement per loss channel}
Here we will derive expressions for the fractions of the emitted power that are absorbed by the antenna, radiated as dipole radiation into the far-field or lost through cavity leakage channels. These expressions are used in \cref{sec:SI-COMSOL} to fit the radiated and absorbed powers extracted from finite-element simulations and thus obtain the necessary antenna and cavity parameters to make a prediction of enhancement in the hybrid system (as shown in \cref{fig:COMSOLComparison}). Also, they are used in \cref{sec:Efficiency} of the main text to calculate the efficiency with which power can be extracted through the cavity decay channel. To know the antenna absorption and the dipolar radiation, we calculate the work done by the absorptive force on the antenna dipole, and by the Abraham-Lorentz force on the total dipole of the source and the antenna, respectively. For power going into the cavity decay channels, we multiply the corresponding decay rate by the energy in the cavity mode.

\subsection{Antenna absorption}
The work done by any force $F$ on a particle moving a distance $dx$ is $F dx$. From the equation of motion \cref{Eq:EOMp1}, we can recognize the 'absorptive' force, i.e. the force describing material absorption, working on the antenna as 
\begin{equation}
F_\mathrm{abs} = \Re{ m \gi \frac{\dot{p}}{q} }.
\end{equation}
The power absorbed in the antenna is the oscillation frequency times the cycle-averaged work done by the absorptive force:
\begin{equation}
P_\mathrm{abs} = \frac{\omega}{2 \pi} \int_0^T F_\mathrm{abs} \frac{dx}{dt} dt
\end{equation}
with $T=\frac{2 \pi}{\omega}$ the cycle time. With $\frac{dx}{dt} = \Re{\frac{\dot{p}}{q}}$ we find
\begin{align}
P_\mathrm{abs} &= \frac{\omega^2}{2 \beta} \gi |p|^2 .
\end{align}
Using \cref{Eq:phyb} for the antenna dipole moment, we arrive at:
\begin{equation}
P_\mathrm{abs} = \frac{\omega^2}{2 \beta} \gi |\ahyb|^2 |\Gbg + \chihom |^2 |\pdr|^2
\end{equation}
To express this power in terms of enhancement, we divide by the homogeneous radiated power (\cref{Eq:Larmor}):
\begin{equation}
\etabs= \frac{6 \pi \eo c^3 }{\oo^2 n} \frac{\gi}{\beta} |\ahyb|^2 |\Gbg + \chihom |^2 \label{Eq:etpabs}
\end{equation}

\subsection{Dipole radiation by antenna and source}
To calculate exactly the power radiated by the source and the antenna, one should calculate the overlap in their radiation patterns by integrating the Poynting flux of their added scattered fields over an enclosing surface. However, to first order we can assume that if the distance $\delta \r$ between source and antenna is sufficiently small (i.e. $\delta \r \ll \lambda $, their radiation patterns overlap entirely. In that case, we may consider them as one effective dipole with total dipole moment $\ptot = \pdr + p$ \cite{Mertens2007}. The force responsible for the radiation of this dipole is the Abraham-Lorentz force. A similar analysis as done for the antenna dissipation then leads to a radiated power by the antenna and source
\begin{align}
P_\mathrm{p,rad} &= \frac{\omega^2}{2 \beta} \gr |\ptot|^2 \\
 &= \frac{\omega^2}{2 \beta} \gr |1+ \ahyb \left(\Gbg + \chihom \right) |^2 |\pdr|^2 .
\end{align}
with $\gr$ the radiative damping rate from \cref{Eq:gr}. The corresponding enhancement then becomes:
\begin{equation}
\etprad = |1+ \ahyb \left(\Gbg + \chihom \right) |^2 . \label{Eq:etprad}
\end{equation}
This answer is intuitive, as it is just the square of the enhancement of the total dipole moment with respect to that of the source.

\subsection{Losses by the cavity}
The cavity damping rate $\kk$ is usually composed of several energy loss terms. These could include e.g. radiation to the far-field, material absorption, scattering by defects or coupling to a waveguide. In the following we will assume the cavity has 2 separate loss channels: radiation and another loss mechanism, which we will assume for now to be dissipation in the cavity. The results can be straightforwardly generalized to include an arbitrary number of loss channels. 
An important assumption is that radiation in these loss channels does not interfere with that in other loss channels or with the far-field dipole radiation by antenna and source. The latter assumption was already required to derive the equations of motion. Interference of radiation in antenna and cavity loss channels would lead to complex coupling rates \cite{Ruesink2015}.

Consider a cavity with a radiation loss rate $\kr$ and a loss rate $\kabs$ due to absorption, such that $\kk =\kr+\kabs$. The power emitted by the cavity into the radiation channel is then:
\begin{align}
P_\mathrm{c,rad} &= \kr \Um \notag \\
&= \frac{\kr }{2 } \eo \ee \Veff \, |\Ec|^2 ,
\end{align}
where we have used \cref{Eq:SI-Um,Eq:SI-Veff} to rewrite the mode energy $U_m$. We can use \cref{Eq:Echyb} for $\Ec$, which leads to:
\begin{equation}
P_\mathrm{c,rad} =  \frac{\kr }{2 } \eo \ee \Veff  \, |\chihyb \left( 1+ \ahom \Gbg \right) |^2 | \pdr |^2.
\end{equation}
Division by the homogeneous radiated power gives the emission enhancement emitted as cavity radiation:
\begin{equation}
\etcrad =  \frac{6 \pi \eo c^3}{\oo^4 n} \kr \eo \ee \Veff  \, |\chihyb \left( 1+ \ahom \Gbg \right) |^2 . \label{Eq:etcrad}
\end{equation}
Similarly, the power that is absorbed in the cavity and the corresponding enhancement are:
\begin{align}
P_\mathrm{c,abs} &=  \frac{\kabs }{2 } \eo \ee \Veff  \, |\chihyb \left( 1+ \ahom \Gbg \right) |^2 | \pdr |^2 , \\
\etcabs &=  \frac{6 \pi \eo c^3}{\oo^4 n} \kabs \eo \ee \Veff  \, |\chihyb \left( 1+ \ahom \Gbg \right) |^2 . \label{Eq:etcabs}
\end{align}

\subsection{Consistency check}
The sum of the enhancements in separate loss channels should match our expressions (\cref{Eq:ettot_hyb,Eq:etptot,Eq:etctot}) for total enhancement. Indeed, we see that this is the case, both for the bare components and for the hybrid system. For a bare cavity, there is perfect agreement. For a bare antenna, a small deviation remains, which can be assigned to the approximation made by assuming a 100\% overlap between source and antenna radiation profiles. The deviation in enhancement is less than 0.5\% of the total enhancement for the antenna-source geometry used in this paper.

\begin{figure*}[htb]
\begin{center}
\includegraphics[width=\textwidth]{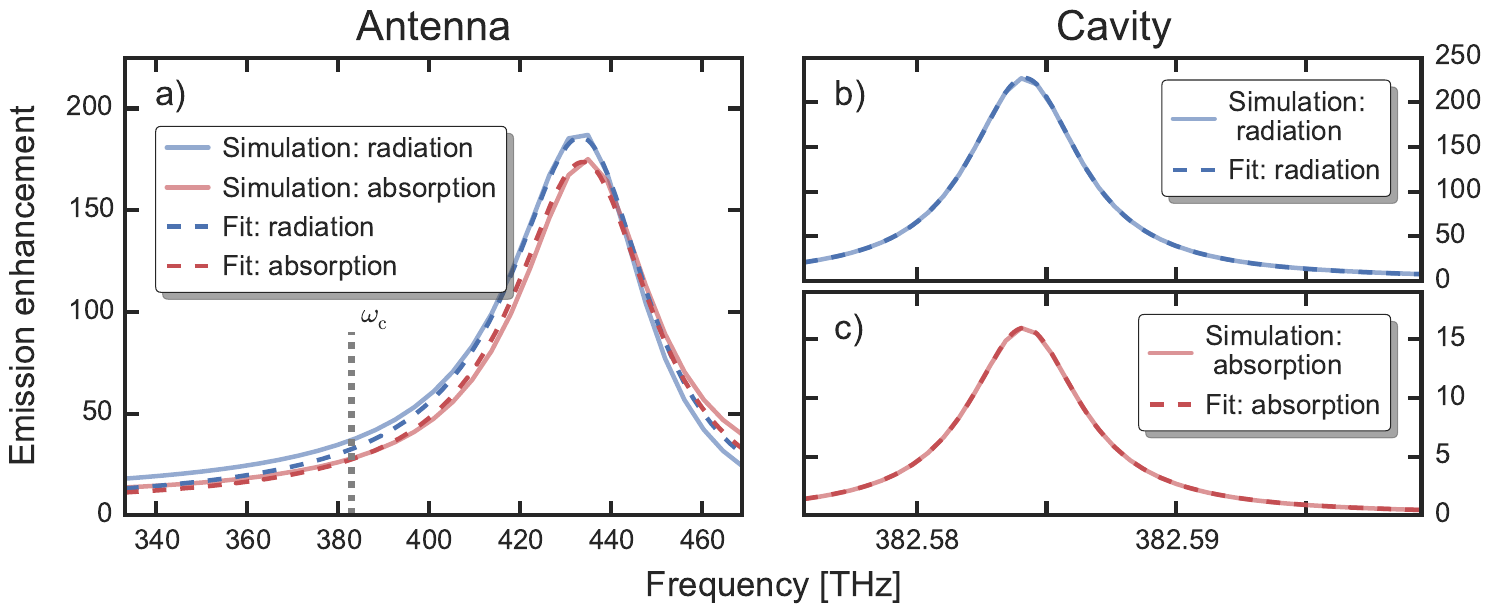} 
\caption{\textbf{a)} Fits to the bare antenna enhancement spectra. The grey dashed line indicates the cavity resonance frequency $\ooc$. The bare antenna was placed 50 nm above an infinite silicon nitride substrate. It can be seen that the fit to the antenna radiation slightly underestimates the radiative enhancement $\ooc$, which explains why the prediction of the oscillator model also underestimates radiative enhancements for the hybrid system, as shown in \cref{fig:COMSOLComparison}. Deviations of the antenna spectra from the lorentzian fits are likely because only a spherical or elipsoidal antenna in vacuum, with metal parameters described by the \emph{unmodified} Drude model, has a strictly lorentzian lineshape. Here we use a modified Drude model \cite{Penninkhof2008}.
\textbf{b)} and \textbf{c)}: Fits to the bare antenna radiative (\textbf{b}) and absorptive (\textbf{c}) enhancement spectra. } \label{fig:Fits} 
\end{center}
\end{figure*}

\section{\label{sec:SI-COMSOL}Finite-element simulations}
In this section we describe the finite-element simulations of the bare cavity and antenna, and of the hybrid system, which were discussed in \cref{sec:COMSOL}. We describe how we retrieved radiative and dissipative enhancements from the simulations, and how we fitted bare component enhancements to obtain the cavity and antenna parameters.

The antenna and cavity geometries are described in \cref{sec:COMSOL} of the main text, and shown in \cref{fig:ModeProfiles}. Cavity spectra are obtained by sweeping the oscillation frequency of a point source placed 50 nm above the disk surface, 300 nm inward from the disk edge (see \cref{fig:ModeProfiles}c) and oriented in the radial direction (in a cylindrical coordinate system with the center of the disk as origin). We integrate the Poynting flux over a surface enclosing the cavity and source, and calculate the absorption in the disk, which are then both normalized to Larmors formula (\cref{Eq:Larmor}) to obtain respectively the radiative and absorptive emission enhancement.

A similar procedure was used for the antenna enhancement spectra, with the source now placed 12 nm from the tip of the antenna (see \cref{fig:ModeProfiles}b) and oriented along the antenna long axis. Radiative and absorptive enhancements were calculated as described above. In the simulation of the hybrid system, the antenna is placed 50 nm above the surface of the disk. The presence of high-index silicon nitride in the antenna near-field red-shifts the antenna resonance frequency slightly (by $\sim$5 THz), which is not captured by the coupled oscillator model. We account for it by simulating the bare antenna 50 nm above an infinite substrate of nitride. 

In \cref{fig:Fits} we show the enhancement spectra obtained from COMSOL simulations for the bare antenna and bare cavity. Also shown are the fits to these spectra. We use \cref{Eq:etpabs} and \cref{Eq:etprad} to fit antenna radiation and dissipation, respectively, with $\oz$, $\gi$, $\beta$ and the antenna-source center-to-center distance $\Delta r$ as fitting parameters. Fitting the cavity radiation and absorption is done using \cref{Eq:etcrad} and \cref{Eq:etcabs}, respectively, with $\ooc$, $\kr$, $\kabs$ and $\Veff$ as fitting parameters. Importantly, for both the cavity and the antenna, the fit routine fits absorption and radiation simultaneously. That is, it calculates the sum of the squared errors between the radiation data and fit, and between the absorption data and fit, and minimizes the sum of the two, using a nonlinear minimization routine. This allows an unambiguous retrieval of the cavity and antenna parameters. 

During the simulation of the hybrid system, the antenna was centered 50 nm above the disk and 218 nm from the edge of the disk. This ensured that the source position with respect to neither disk nor antenna was changed with respect to the simulations of the bare components. The antenna and source positions above the disk were chosen such that cavity mode intensity was approximately equal (within 12\%) at the source and the antenna positions, as was assumed in the oscillator model (see \cref{sec:SI-TotalEnh-DipoleDrive}).

\bibliography{CHOpaper,Tame2013ref}
\end{document}